\pdfoutput=1
\pdfoutput=1
\documentclass[acmtog]{acmart}
\acmSubmissionID{1106}
\AtBeginDocument{%
  \providecommand\BibTeX{{%
    \normalfont B\kern-0.5em{\scshape i\kern-0.25em b}\kern-0.8em\TeX}}}

\definecolor{MyRed}{rgb}{0.65,0.07,0.09}

\definecolor{MyGreen}{rgb}{0.18,0.55,0.09}

\usepackage{wrapfig}
\usepackage{diagbox}
\usepackage{verbatim}
\usepackage{amssymb}
\usepackage{latexsym}
\usepackage{lineno}
\usepackage{xspace}
\usepackage{color}
\graphicspath{{figures/}}
\usepackage{amsmath,bm}
\usepackage{psfrag}
\usepackage{mathtools}

\theoremstyle{definition}

\usepackage{multirow, tabularx}
\newcolumntype{Y}{>{\centering\arraybackslash}X}
\usepackage{capt-of}%
\usepackage{array, boldline, makecell, booktabs}

\usepackage{array}
\usepackage{pifont}
\usepackage{psfrag}
\usepackage{makecell}
\usepackage{algorithm}
\usepackage{booktabs}
\usepackage{graphicx}
\usepackage{subcaption}
\usepackage{stfloats} 
\usepackage{appendix}
\usepackage[noend]{algpseudocode}

\usepackage{xcolor}
\usepackage{graphicx}

\usepackage{multirow}

\newcommand{\bo}[1]{\textcolor{blue}{[Bo: #1]}}

\newif\ifverbose
\verbosetrue

\ifverbose

\newcommand{\vb}[1]{\textcolor{red}{#1}}
\else

\newcommand{\vb}[1]{}
\fi

\usepackage{url}
\usepackage{xcolor}
\definecolor{newcolor}{rgb}{.8,.349,.1}

\newcommand{\duowen}[1]{\textcolor{black}{#1}}

\copyrightyear{2024}
\acmYear{2024}
\setcopyright{rightsretained}
\acmConference[SIGGRAPH Conference Papers '24]{Special Interest Group on Computer Graphics and Interactive Techniques Conference Conference Papers '24}{July 27-August 1, 2024}{Denver, CO, USA}
\acmBooktitle{Special Interest Group on Computer Graphics and Interactive Techniques Conference Conference Papers '24 (SIGGRAPH Conference Papers '24), July 27-August 1, 2024, Denver, CO, USA}
\acmDOI{10.1145/3641519.3657514}
\acmISBN{979-8-4007-0525-0/24/07}

\begin{document}
\author{Zhiqi Li}
\affiliation{
\institution{Georgia Institute of Technology}
\country{USA}
}\email{ zli3167@gatech.edu}

\author{Barnab\'{a}s B\"{o}rcs\"{o}k}
\affiliation{
\institution{Georgia Institute of Technology}
\country{USA}
}\email{borcsok@gatech.edu}

\author{Duowen Chen}
\affiliation{
\institution{Georgia Institute of Technology}
\country{USA}
}\email{dchen322@gatech.edu}

\author{Yutong Sun}
\affiliation{
\institution{Georgia Institute of Technology}
\country{USA}
}\email{ysun730@gatech.edu}

\author{Bo Zhu}
\affiliation{
\institution{Georgia Institute of Technology}
\country{USA}
}\email{bo.zhu@gatech.edu}

\author{Greg Turk}
\affiliation{
\institution{Georgia Institute of Technology}
\country{USA}
}\email{turk@cc.gatech.edu}


\title{Lagrangian Covector Fluid with Free Surface}

\begin{abstract}
This paper introduces a novel Lagrangian fluid solver based on covector flow maps. We aim to address the challenges of establishing a robust flow-map solver for incompressible fluids under complex boundary conditions. 
Our key idea is to use particle trajectories to establish precise flow maps and tailor path integrals of physical quantities along these trajectories to reformulate the Poisson problem during the projection step. We devise a decoupling mechanism based on path-integral identities from flow-map theory. This mechanism integrates long-range flow maps for the main fluid body into a short-range projection framework, ensuring a robust treatment of free boundaries. We show that our method can effectively transform a long-range projection problem with integral boundaries into a Poisson problem with standard boundary conditions --- specifically, zero Dirichlet on the free surface and zero Neumann on solid boundaries. This transformation significantly enhances robustness and accuracy, extending the applicability of flow-map methods to complex free-surface problems. 
\end{abstract}



\begin{teaserfigure}
 \centering
 \includegraphics[width=.38467\textwidth]{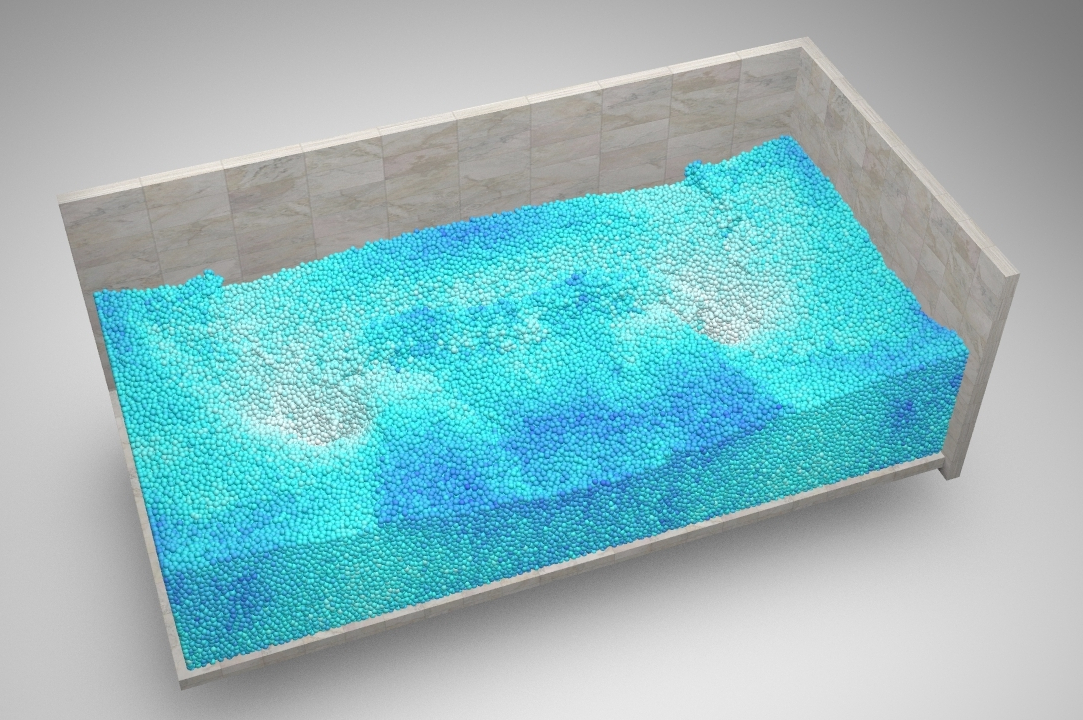}
 \includegraphics[width=.34103\textwidth]{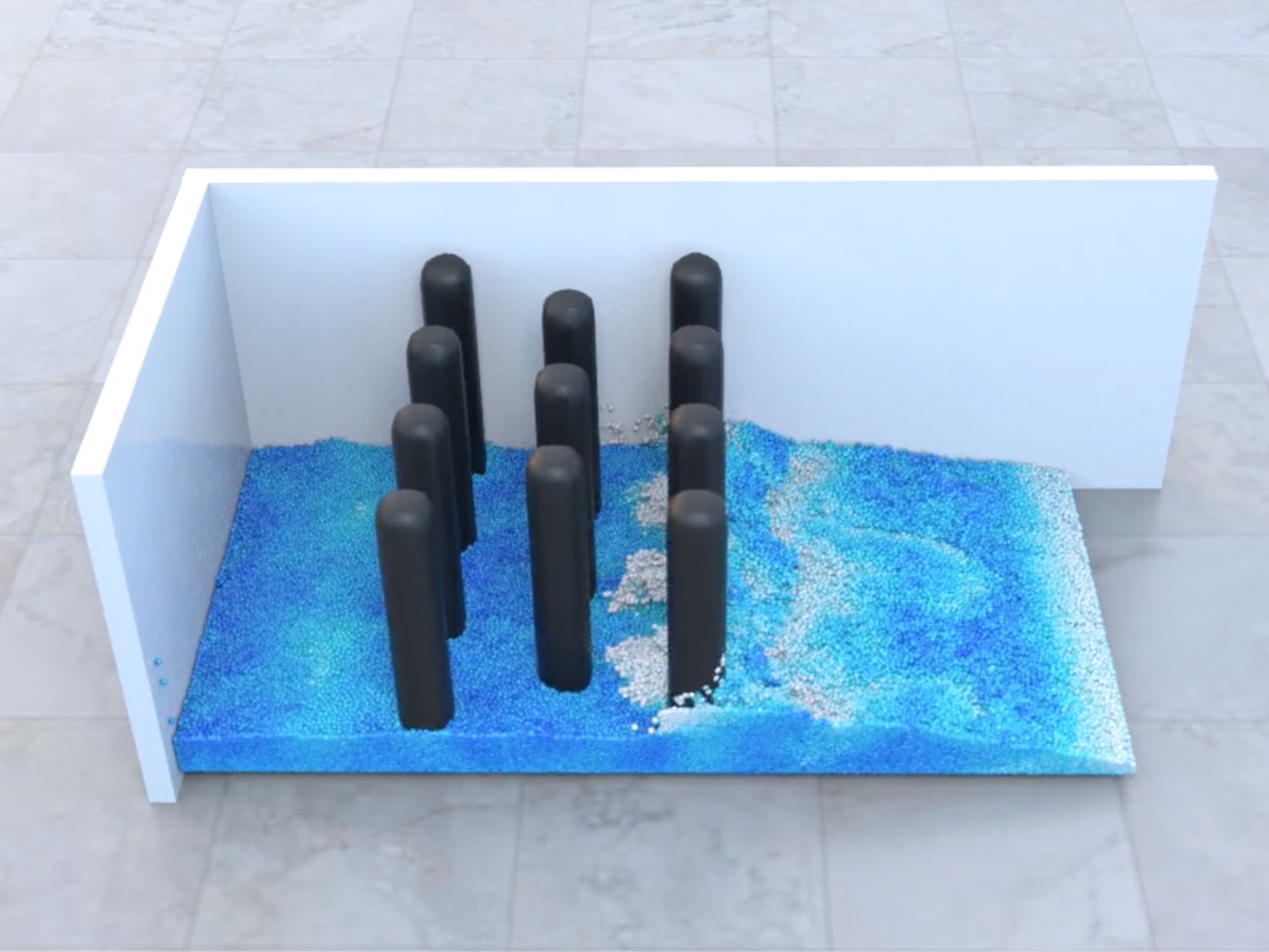}
 \includegraphics[width=.265\textwidth]{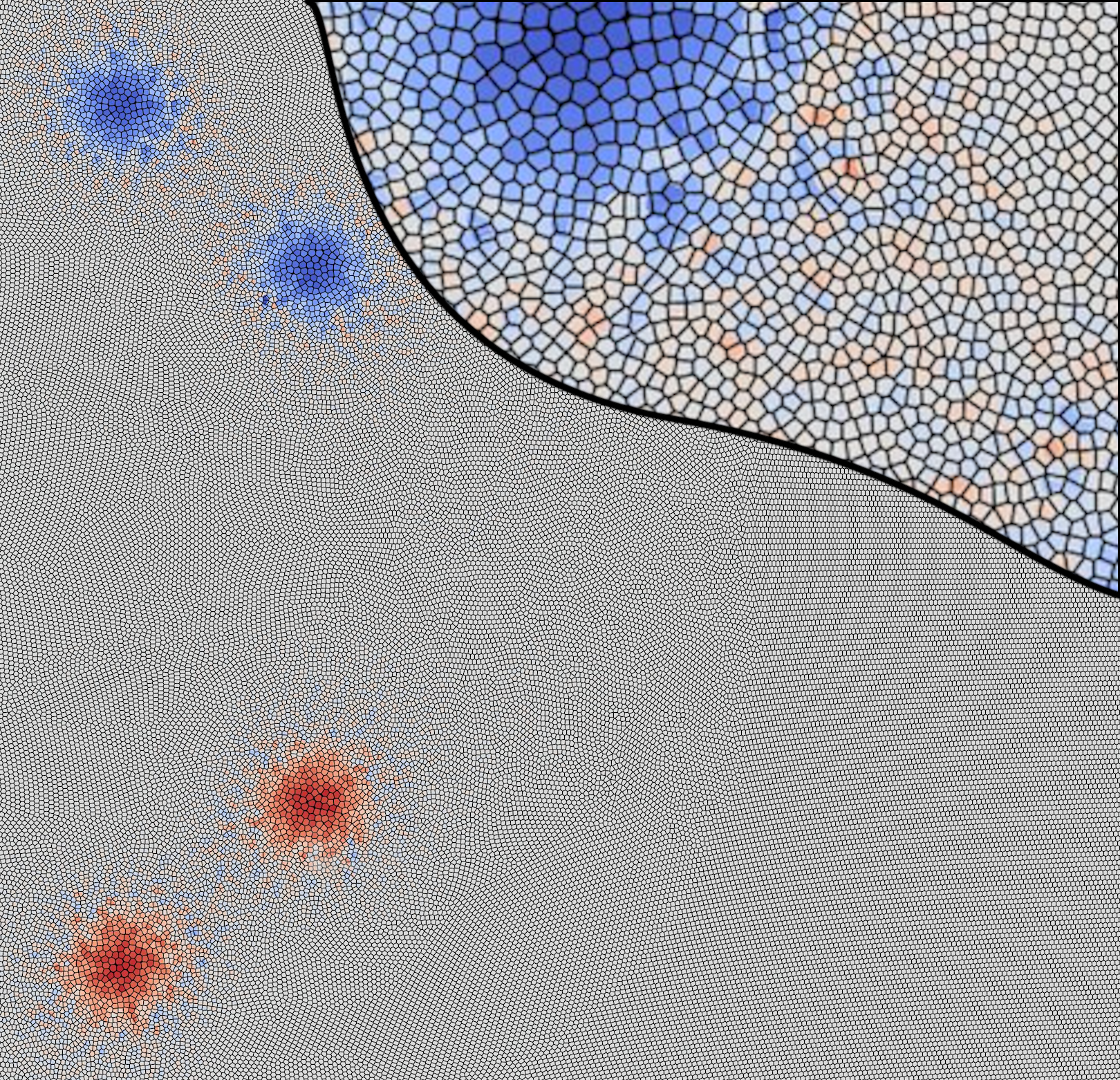}
 \caption{Examples of our method handling demanding fluid simulation scenarios. In three dimensions, a two-sink setup draining a tank of water \textit{(left)}, and generated waves hitting multiple solid objects \textit{(center)}. Our method achieves state-of-the-art performance amongst purely particle-based methods for simulating dynamic vortical structures, such as two-dimensional leapfrogging \textit{(right)}.}
 \label{fig:teaser}
\end{teaserfigure}

\maketitle

\section{Introduction}

Flow map methods have garnered increasing interest in both computational physics and computer graphics communities in recent years, as evidenced by the emergence of covector/impulse-based \cite{nabizadeh2022covector,deng2023fluid} and vorticity-based \cite{mercier2020characteristic, yin2021characteristic, yin2023characteristic} solvers that are known for their exceptional preservation of vortical structures. The key to developing a flow-map method is constructing an efficient and accurate representation that maps any given point from the initial to the current frame (and vice versa if a bi-directional mapping process is necessary). This flow map, or the relationship between the two endpoints of every mapping trajectory sample, was previously realized by advecting the spatial coordinates \cite{hachisuka2005combined, sato2017long, sato2018spatially, tessendorf2015advection, deng2023fluid}, with improved accuracy later achieved by tracing particles backward over a recorded spatiotemporal velocity field, both of which were implemented in a fixed, Eulerian domain.

One of the main challenges in the impulse/covector fluid models with their flow-map-based implementations is addressing free-surface boundaries. Existing approaches struggle to simulate free-surface fluids 
due to inherent difficulties with free-surface boundary conditions that are impractical to manage using traditional numerical solvers. In standard free-surface solvers built on velocity space, fluid incompressibility is enforced by solving a Poisson problem with zero Dirichlet boundary conditions on the free surface (assuming zero air pressure) and appropriate Neumann boundary conditions on solid boundaries. In contrast, the free-surface boundary conditions for a covector flow map model pose greater challenges. This difficulty arises from the complexity of calculating the kinetic energy integral on the free boundary over the entire flow map interval. 

In typical liquid simulation scenarios in computer graphics, the fluid surface undergoes extensive geometric and topological transitions over time. Consequently, a fluid particle may appear on the surface at a certain instant and then merge into the fluid body later. This dynamics makes it impractical to track these particles' statuses consistently and determine whether they are on the free surface at any given time over a high-dimensional spatiotemporal space. However, the accuracy of the boundary condition depends on the path integral of \textit{all particles currently on the free surface over the entire flow-map period}. Due to these challenges with traditional numerical solutions, the covector/impulse frameworks and their flow-map implementation are confined to solving fluid without a free surface and can produce smoke animations only. 

This paper makes the first step toward addressing the free-surface boundary challenges for covector flow-map methods. We attack the problem from the Lagrangian perspective by treating each flow-map sample as a Lagrangian particle trajectory. Under this Lagrangian view, we proposed a novel mathematical framework based on the path integral identities in flow-map theory to decouple the mapping and projection steps in a conventional covector flow-map algorithm. Our key idea is to leverage these mathematical identities to flexibly control the integral intervals for different physical quantities along each Lagrangian path, and then transform the boundary conditions from the long-range integral (through all flow map time steps) to a \color{black}short-range \color{black}integral (within a single time step), and eventually to a standard zero Dirichlet boundary to fit into the existing Poisson solver. By doing so, we can decouple the long-range map for velocity, which serves as the enabling mechanism for the vortical expressiveness of flow-map methods, and the pressure projection step, which is fundamental for simulating incompressible flow, without any model degeneration or artificial numerical blending.

Our proposed Lagrangian covector solver comprises three components: a Lagrangian model for long-range flow maps and path integrals, a reformulated Poisson solver designed for handling complex boundaries, and a Voronoi-based numerical implementation. These components synergistically establish a particle-based framework that facilitates integral-flexible flow maps for the first covector solver capable of handling free surfaces. 


We summarize our main contributions as follows:
\begin{itemize}
    \item We proposed a novel Lagrangian flow map model characterizing the path integral form of covector fluid.
    \item We proposed a novel mechanism to incorporate long-range flow maps into a short-range projection step and reformulated it into a standard Poisson problem.
    \item We proposed the first free-surface covector fluid model based on a Voronoi implementation. 
\end{itemize}

 
\section{Related Work}


\paragraph{Particle-based fluid simulation}
\cite{de2015power} introduced power particles to simulate incompressible fluids. Their geometrically inspired method offers precise pressure solving and an even distribution of particles by describing the fluid motion as a series of well-shaped power diagrams. \cite{zhai2018fluid} accelerated the construction of power diagrams on GPUs, adopting ghost particles for fluid-air interactions, while \cite{levy2022partial} introduced a more precise technique to calculate free-surface interactions, framing it as an optimal transport problem.

%
\paragraph{Flow-map methods}
The method of characteristic mapping (MCM) of \citet{wiggert1976numerical} was first introduced to the graphics community by \citet{tessendorf2011characteristic}. Due to its superiority in dealing with numerical dissipation through long-range mapping, some methods \cite{hachisuka2005combined, sato2017long, sato2018spatially, tessendorf2015advection} trade off computational cost for better accuracy utilizing virtual particles for tracking the mapping. Subsequently, \citet{nabizadeh2022covector} combined this with an impulse fluid model \cite{cortez1996impulse} and \citet{qu2019efficient} developed a bidirectional mapping to prevent dissipation better. \citet{deng2023fluid} stores intermediate velocity fields using a neural network for storage compression. We track such mapping directly with particles. 
\color{black}
Compared to previous flow map methods (e.g., \cite{sato2017long}), our method relies on calculating path integrals on particles directly without any backtrace step, eliminating any extra velocity buffer. 
\color{black}


\paragraph{Gauge-based fluid}
The concept of the \textit{impulse variable} was initially introduced in \cite{buttke1992lagrangian}, reformulating the incompressible Navier-Stokes Equation with a gauge variable and transformation \cite{oseledets1989new, roberts1972hamiltonian, buttke1993velicity}. Various gauges have been proposed for applications like boundary treatment, numerical stability, and turbulence simulation \cite{buttke1993velicity, buttke1993turbulence, weinan2003gauge, cortez1996impulse, summers2000representation}. \citet{saye2016interfacial} and \citet{saye2017implicit} employed a different gauge for interfacial discontinuity issues. In the graphics community, researchers like \citet{feng2022impulse}, \citet{Xiong2022Clebsch}, \citet{yang2021clebsch}, and \citet{nabizadeh2022covector} have explored this area, though facing challenges with accurate advection. Flow maps, as shown in \cite{deng2023fluid}, are effective for advecting such variables, and this method has been adapted to particles in our research.


\section{Mathematical Foundation}

\newcolumntype{z}{X}
\newcolumntype{s}{>{\hsize=.25\hsize}X}
\begin{table}[t]
\centering\small
\begin{tabularx}{0.47\textwidth}{scz}
\hlineB{3}
Notation & Type & Definition\\
\hlineB{2.5}
\hspace{12pt}$r$ & scalar & Current time\\
\hlineB{1}
\hspace{12pt}$s$ & scalar & Initial time\\
\hlineB{1}
\hspace{12pt}$s'$ & scalar & One time step before the current time\\
\hlineB{1}
\hspace{12pt}$\Phi_a^b$ & vector & Forward flow map from time $a$ to $b$\\
\hlineB{1}
\hspace{12pt}$\Psi_b^a$ & vector & Backward flow map from time $b$ to $a$\\
\hlineB{1}
\hspace{12pt}$\mathcal{F}_a^b$ & matrix & Jacobian of forward flow map $\Phi_a^b$\\
\hlineB{1}
\hspace{12pt}$\mathcal{T}_b^a$ & matrix & Jacobian of backward flow map $\Psi_b^a$\\
\hlineB{1}
\hspace{12pt}$p_t$ & scalar & Pressure at time $t$\\
\hlineB{1}
\hspace{12pt}$\lambda_t$ & scalar & Lagrangian pressure calculated at time $t$\\
\hlineB{1}
\hspace{12pt}$\Lambda_a^b$ & scalar & Integration of $\lambda$ from time $a$ to  $b$\\
\hlineB{1}
\hspace{12pt}$\mathbf{u}_{a\to b}^M$ & vector & \textbf{Mapped} velocity at time $b$ by covector flow map from time $a$\\
\hlineB{1}
\hspace{12pt}$\mathbf{u}_{a\to b}^A$ & vector & \textbf{Advected} velocity at time $b$ by particle velocity from time $a$\\
\hlineB{3}
\end{tabularx}
\vspace{5pt}
\caption{Summary of important notations used in the paper.}
\label{tab: notation_table}
\vspace{-0.45in}  
\end{table}

\paragraph{Flow Maps} 
\begin{wrapfigure}[11]{r}{0.3\textwidth}
    \centering
    \vspace{-0.2cm}
    \includegraphics[width=0.3\textwidth]{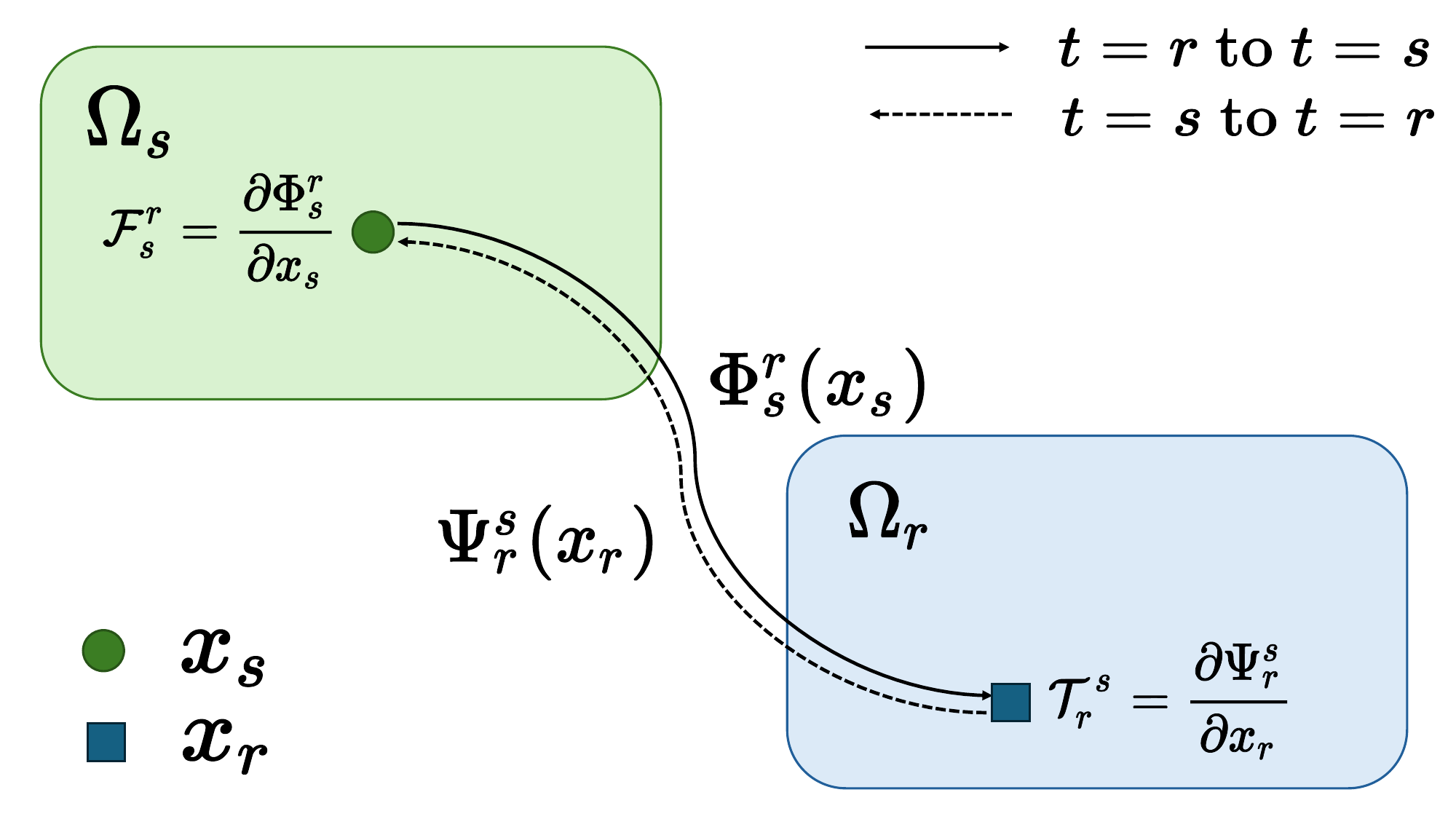}
    \vspace{-0.75cm}\caption{Forward and backward flow maps defined on a particle.}\label{fig:fm}
\end{wrapfigure}
To define the flow map, we start with an initial domain \(\Omega_s\) at time \(s\) and its current domain \(\Omega_r\) at time \(r\), with \(s < r\). A particle at \(\mathbf{x}_s \in \Omega_s\) at time \(s\) moves to \(\mathbf{x}_r \in \Omega_r\) at time \(r\) by velocity \(\mathbf{u}_t,t\in[s,r]\). The relationship between \(\mathbf{x}_s\) and \(\mathbf{x}_r\) is defined through two flow map functions, including the forward flow map \(\mathbf{x}_r = \Phi_s^r(\mathbf{x}_s)\) and the backward flow map \(\mathbf{x}_s = \bm \Psi_r^s(\mathbf{x}_r)\), with Jacobians \(\mathcal{F}_s^r = \frac{\partial \Phi_s^r}{\partial \mathbf{x}_s}\) and \(\mathcal{T}_r^s = \frac{\partial \Psi_r^s}{\partial \mathbf{x}_r}\).
These two flow maps satisfy the equations \color{black}\(\Phi_s^r \circ \Psi_r^s = \text{id}_s\)\color{black} and \color{black}\(\Psi_s^r \circ \Phi_r^s = \text{id}_r\)\color{black}, where \(\text{id}_s\) and \(\text{id}_r\) denote the identity transformations in domains \(\Omega_s\) and \(\Omega_r\) respectively. 

Considering the domain \(\Omega_t \in \mathbb{R}^n\) (where \(n = 2, 3, \ldots\)) at any time \(t\) as a linear vector space, the flow maps \(\Phi_s^r: \Omega_s \to \Omega_r\) and \(\Psi_r^s: \Omega_r \to \Omega_s\) are mappings between these linear spaces. 
Given a scalar field \(q_s\) at time \(s\), a scalar field ${\Psi_t^s}^*q_t$ on \(\Omega_t\) can be induced via the map $({\Psi_t^s}^*q_s)(\mathbf{\mathbf{x}}) \overset{\Delta}{=} q_s(\Psi_t^s(\mathbf{x}))$ (\(^*\) means induction, where a map ${\Psi_t^s}^*$ between fields is induced by a map $\Psi_t^s$ between domains). This induced scalar field is called the pullback of \(q_s\) by the mapping \(\Psi_t^s(\mathbf{x})\) (the pullback by \(\Phi_s^t(\mathbf{x})\) is similar).



\paragraph{Covector Preliminaries} In the domain \(\Omega_t\) at any time \(t\), regarded as a linear space, we can define tangent spaces \(T_\mathbf{x} \Omega_t\) and cotangent spaces \(T^*_\mathbf{x} \Omega_t\) at any point \(\mathbf{x}\) within \(\Omega_t\). For \(\Omega_t \subset \mathbb{R}^n\), the tangent spaces \(T_\mathbf{x} \Omega_t = \mathbb{R}^n\) encompass all vectors originating from \(\mathbf{x}\), while the cotangent space \(T^*_\mathbf{x} \Omega_t\) comprises all linear functions (termed cotangent vectors or covectors) defined on \(T_\mathbf{x} \Omega_t\). Given an inner product in \(\Omega_t \subset \mathbb{R}^n\), any tangent vector \(\mathbf{v}_\mathbf{x} \in T_\mathbf{x} \Omega_t\) induces a covector in \(T^*_\mathbf{x} \Omega_t\), such that \(\mathbf{v}_\mathbf{x}^\flat(\mathbf{w}) = \langle \mathbf{v}, \mathbf{w} \rangle\) for all \(\mathbf{w} \in T_\mathbf{x} \Omega_t\). Here, \(\flat\) denotes the conversion of a tangent vector to a covector, and \(\sharp\) signifies the reverse conversion. 
A vector field \(\mathbf{v}(\mathbf{x})\) (or covector field \(\mathbf{v}^\flat(\mathbf{x})\)) is formed by selecting one vector \(\color{black}\mathbf{v}_\mathbf{x} = \mathbf{v}(\mathbf{x})\color{black}\) (or one covector \(\mathbf{v}_\mathbf{x}^\flat = \mathbf{v}^\flat(\mathbf{x})\)) from each tangent space \(T_\mathbf{x} \Omega_t\) (or the cotangent space \(T^*_\mathbf{x} \Omega_t\)) at every point \(\mathbf{x}\). \color{black}The gradient \(d q\) of a scalar field \(q\) can be considered as a vector field. \color{black}Refer to \cite{Crane2013Exteriorcalculus} for details.  

\paragraph{Lie Advection} For any covector field $\color{black}\mathbf{v}_s^\flat\color{black}$ taken from $\Omega_s$, $\Psi_t^s$ induces a covector field ${\Psi_t^s}^*\color{black}\mathbf{v}_s^\flat\color{black}$ on $\Omega_t$ as $({\Psi_t^s}^*\color{black}\mathbf{v}_s^\flat\color{black})(\mathbf{x}) \overset{\Delta}{=}{\color{black}[ {\mathcal{T}_t^s}^T(\mathbf{x})\mathbf{v}_s(\Psi_t^s(\mathbf{x}))]}^\flat\color{black}$,
which is referred to as the pullback of the covector field $\color{black}\mathbf{v}_s^\flat\color{black}$ by $\Psi_t^s$. The pullback of $\color{black}\mathbf{v}_s^\flat\color{black}$ at any time $t$ satisfies an Lie advection equation:$\left(\frac{\partial}{\partial t} + \mathcal{L}_\mathbf{u}\right)\color{black}\mathbf{v}_t^\flat\color{black} = 0$, where $\mathcal{L}_\mathbf{u}$ is the Lie derivative, and in $\Omega_r \subset \mathbb{R}^n$, it is given by $\color{black}(\mathcal{L}_\mathbf{u} \mathbf{v}_t^\flat)^\sharp = (\mathbf{u} \cdot \nabla) \mathbf{v}_t + (\nabla \mathbf{u})^T \cdot \mathbf{v}_t\color{black}$. Therefore if there is a vector field that satisfies this advection equation, at any time $t$, $\color{black}\mathbf{v}^\flat_t\color{black}$ can be described through the pullback of the initial $\color{black}\mathbf{v}^\flat_s\color{black}$ through pullback.  

\section{Physical Model}

\paragraph{Covector Incompressible Fluid}
We model incompressible flow using Euler equations by assuming viscosity zero and density one:
\begin{align}
\begin{dcases}
    \frac{\partial \mathbf{u}}{\partial t}+\color{black}(\mathbf{u} \cdot \nabla) \mathbf{u}\color{black} + \nabla p = 0, \\ 
    \nabla \cdot \mathbf{u} = 0,
    \label{equ:euler}
\end{dcases}
\end{align}
with \(\mathbf{u}\) and \(p\) specifying the fluid velocity and pressure. The first equation describes the momentum conservation, and the second equation specifies incompressibility. 
According to the covector model proposed in \cite{nabizadeh2022covector}, Equation~\ref{equ:euler} can be reformulated into its covector form as \color{black}$\frac{\partial \mathbf{u}}{\partial t}+(\mathbf{u} \cdot \nabla) \mathbf{u}+\nabla \mathbf{u}^T\cdot \mathbf{u}+ \nabla (p-\frac{1}{2}|\mathbf{u}|^2)=0$\color{black}, which can be further written with Lie advection as 
\begin{equation}
    (\frac{\partial}{\partial t}+\mathcal{L}_\mathbf{u})\mathbf{u}^\flat+d (p-\frac{1}{2}|\mathbf{u}|^2):=(\frac{\partial}{\partial t}+\mathcal{L}_\mathbf{u})\mathbf{u}^\flat+\color{black}d \lambda\color{black}=0,
    \label{equ:euler_covector_2}
\end{equation}
where $\lambda=p-\frac{1}{2}|\mathbf{u}|^2$ is defined as Lagrangian pressure. 

\paragraph{Covector Fluid on Flow Maps}
The solution \(\mathbf{u}_r\) of Equation \ref{equ:euler_covector_2} at $r$ includes terms obtained by pulling back the velocity field as a covector through the flow map. This process is illustrated by integrating both sides of Equation \ref{equ:euler_covector_2} from time \(s\) to \(r\), represented as a covector: 
 \begin{equation}
    \begin{aligned}
    \mathbf{u}_r^\flat &= {\Psi_r^s}^*\mathbf{u}_s^\flat - \int_s^r (\color{black}{\Phi_s^\tau \circ \Psi_r^s}\color{black})^* \color{black}d \lambda_\tau\color{black} d\tau, \\
    &= {\Psi_r^s}^*\mathbf{u}_s^\flat - d \int_s^r (\color{black}{\Phi_s^\tau \circ \Psi_r^s}\color{black})^* \lambda_\tau d\tau,
    \end{aligned}
    \label{equ:u_fm_covector}
 \end{equation}
with the second line arising from the commutativity between the differential operator \(\nabla\) (more accurately denoted as ``d'') and the pullback operator (Equation 7 in \cite{nabizadeh2022covector}), and commutativity between \(\nabla\) and the integral sign.

%
Next, we can convert the covector forms and their pullbacks in \autoref{equ:u_fm_covector} into their vector forms as: 
\begin{equation}
\mathbf{u}(\mathbf{x},r) = \underbrace{{\mathcal{T}_r^s}^T \mathbf{u}_s(\Psi_{r}^s(\mathbf{x}),s)}_{\text{mapping}} - \underbrace{\nabla \int_s^r\lambda((\color{black}{\Phi_s^\tau \circ \Psi_r^s})(\mathbf{x}\color{black}),\tau)d\tau}_{\text{projection}},
\label{eq:u_fm}
\end{equation}
which constitutes two main steps to obtain velocity at \((\mathbf{x}, r)\) using a \textit{mapping-projection} scheme. Initially, the mapping step calculates \({\mathcal{T}_r^s}^T \mathbf{u}_s(\color{black}\Psi_{r}^s(\mathbf{x})\color{black}, s)\) as the pullback of the velocity field \(\mathbf{u}_s\) by the flow map. In the projection step, the gradient of \(\int_s^t\lambda((\color{black}{\Phi_s^\tau \circ \Psi_r^s}\color{black})(\mathbf{x}), \tau)d\tau\) is projected from the mapped velocity to enforce incompressibility. 

\section{Path Integral on Lagrangian Flow Maps}
\autoref{eq:u_fm} can be written into a path integral form on a Lagrangian trajectory. For a fluid particle $q$ with position $\mathbf{x}_q(t)$ at any time $t$, the projection term in \autoref{eq:u_fm} is the path integral of $\lambda$ along its trajectory from time $s$ to $r$, which is denoted as 
\begin{equation}
\Lambda_{s,q}^{r}\overset{\Delta}{=}\int_s^r\lambda((\color{black}{\Phi_s^\tau \circ \Psi_r^s}\color{black})(\mathbf{x}_q(r)),\tau)d\tau.    
\end{equation}
Here the subscript \(q\) indicates that this quantity is carried by the particle \(q\).  
For the mapping part, $\mathbf{u}_s(\Psi_{r}^s(\mathbf{\mathbf{x}_q(r)}),s)$ is the velocity of particle $q$ at time $s$, and we simplify the notation as $\mathbf{u}_{s,q}$. 
Because \({\mathcal{T}_r^s}=\frac{\partial \mathbf{x}_s}{\partial \mathbf{x}_r}\) is determined by the positions of all particles in the flow field at time \(r\) and \(s\), it can be carried on fluid particles, denoted as \({\mathcal{T}_{r,q}^s}\).  We obtain the reformulated \autoref{eq:u_fm} as a path integral:
\begin{equation}
\boxed{\mathbf{u}_{r,q} = \underbrace{{\mathcal{T}_{r,q}^s}^T \mathbf{u}_{s,q}}_{\text{mapping}} - \underbrace{\nabla \Lambda_{s,q}^r}_{\text{projection}}}.
\label{eq:u_fm_particle}
\end{equation}

Equation~\ref{eq:u_fm_particle} forms the mathematical foundation of our Lagrangian covector fluid model. This equation, in contrast to Equation~\ref{eq:u_fm}, considers the mapping and projection processes occurring on a particle as it moves along its trajectory in the flow field, which simplifies the formulation by eliminating the need for back-and-forth mappings to identify a Lagrangian point. Thus, we reinterpret the mapping-projection process on a moving particle $q$ as follows: (1) calculate the mapped velocity from the initial time $s$ as $\bm u^M_{s\to r,q}={\mathcal{T}_{r,q}^s}^T \mathbf{u}_{s,q}$, and (2) remove its irrotational component by subtracting the gradient of the path integral of the Lagrangian pressure $\nabla \Lambda_{s,q}^r$.

\section{Incompressibility}
\subsection{Long-Range Mapping, Long-Range Projection}
To determine the velocity of a particle \( q \) at time \( r \), we initially consider a straightforward method for solving Equation~\ref{eq:u_fm_particle}. We present the method as a standard \textit{mapping-projection} scheme:
\begin{enumerate}
    \item (\textbf{Long-Range Mapping}) Calculate the long-range mapped velocity as: \({\mathbf{u}_{s\to r,q}^{M}} = {\mathcal{T}_{r,q}^s}^T \mathbf{u}_{s,q}\);
    \item (\textbf{Long-Range Projection}) Solving the following Poisson equation to obtain $\Lambda_{s,q}^r$:
    \begin{equation}
        \begin{cases}
            \nabla\cdot\nabla \Lambda_{s}^r =\nabla\cdot {\mathbf{u}_{s\to r}^{M}}, & \mathbf{x}\in \Omega\\
            \mathbf{u}_r=\mathbf{u}_b, & \mathbf{x}\in \partial_s \Omega\\
            p=0, & \mathbf{x}\in \partial_f \Omega
        \end{cases}
    \label{eq:possion_1}
    \end{equation}
    where $\mathbf{u}_b$ denotes the velocity of the solid boundary, and \(\partial_s \Omega \) and \(\partial_f \Omega \) denote solid boundary and free surface boundary respectively.
    Then, we carry out a projection as \(\mathbf{u}_{r,q} = {\mathbf{u}_{s\to r,q}^{M}} - \nabla \Lambda_{s,q}^r\) to ensure \(\mathbf{u}_{r}\) satisfies \(\nabla \cdot \mathbf{u}_r = 0\).    
\end{enumerate}
Here, $\mathbf{u}_{s\to r,q}^{M}$ is long-range mapped from the initial time $s$ and is projected to $\mathbf{u}_r$ by a gradient of long-range path integral of pressure $\Lambda_{s,q}^r$ from time $s$ to $t$. 
Hence, the Poisson equation we solve in \autoref{eq:possion_1} leads to a long-range projection of the rotational component accumulated from time $s$ to time $r$. Given these, we refer to this strategy as \textbf{L}ong-Range \textbf{M}apping \textbf{L}ong-Range \textbf{P}rojection (\textbf{LMLP}).  

LMLP has two main issues regarding boundary and performance:
\begin{enumerate}
    \item Setting boundary conditions is challenging. Specifically, at the free boundary \(\partial_f \Omega\), to enforce the boundary condition \(p=0\), we define \(\Lambda_{s,q}^r=\int_s^rp_{\tau,q}-\frac{1}{2}|\mathbf{u}_{\tau,q}|^2 d\tau\). This implies the necessity of establishing a non-zero Dirichlet boundary condition such that \(\Lambda_{s,q}^r=\int_s^r-\frac{1}{2}|\mathbf{u}_{\tau,q}|^2 d\tau\). This condition requires a comprehensive path integral of the kinematic energy across a particle's trajectory. Although seemingly feasible from a Lagrangian perspective, it is crucial to recognize that this path integral only constitutes a valid Dirichlet boundary under the specific condition that \textit{the particle remains on the free surface from time \( s \) to \( t \)}. However, in practical scenarios, this condition is rarely met due to the dynamic topological changes of the surface over time. In the interval from time \( s \) to \( r \), particles may transition between the interior and the surface, rendering the calculation of this integral impractical.
    
    \item The performance issue is notable. The term \({\mathbf{u}_{s\to r,q}^{M}}\) encompasses a substantial divergent component \(\nabla\Lambda_{s,q}^r\), which includes an integral extending from time \(s\) to \(t\). Attempting to eliminate this component through a single projection results in significant computational expenses, primarily due to the increased number of iterations required for solving the Poisson equation.  
\end{enumerate}

\subsection{Short-Range Mapping, Short-Range Projection}
To address the two issues mentioned above, we devise a short-range approach. By setting the flow map's start point to be \textit{only one timestep before $r$, denoted as $s'$}, we can calculate the velocity at time $r$ with the following two steps: 
\begin{enumerate}
    \item (\textbf{Short-Range Mapping}) Calculate the mapped velocity: \(\mathbf{u}_{s'\to r,q}^M = {\mathcal{T}_{r,q}^{s'}}^T \mathbf{u}_{s',q}\)
    \item (\textbf{Short-Range Projection}) Solve the following Poisson equation to obtain $\hat \lambda_q$:
    \begin{equation}
    \begin{cases}
    \nabla\cdot\nabla \hat \lambda =\nabla\cdot {\mathbf{u}_{s'\to r}^{M}}, &  \mathbf{x}\in \Omega\\
     \mathbf{u}_r=\mathbf{u}_b, &  \mathbf{x}\in \partial_s \Omega\\
     p=0 \Rightarrow \hat \lambda = -\frac{1}{2}|\mathbf{u}_{r}^2|\Delta t, & \mathbf{x}\in \partial_f \Omega
     \end{cases}
    \label{eq:possion_2}
    \end{equation}
    where $\hat \lambda_q=\lambda_{r,q}\Delta t$ is the numerical calculation of one step integration $\Lambda_{s',q}^r$. Then, we conduct the divergence-free projection as $\mathbf{u}_{r,q}=\mathbf{u}_{s'\to r,q}^M-\nabla \hat\lambda_q$.
\end{enumerate}
Given both mapping and projection are limited within a short interval from time $s'$ to $r$, we refer the scheme as \color{black}\textbf{S}hort-Range \textbf{M}apping \textbf{S}hort-Range \textbf{P}rojection \color{black}(\textbf{SMSP}).  
\color{black}It is worth noting that SMSP is akin to the Semi-Lagrangian advection schme, as they both involve computing a one-step mapping. The difference lies in the fact that SMSP employs a mapping form based on the Lie advection equation \(\mathbf{u}_{s'\to r,q}^M = {\mathcal{T}_{r,q}^{s'}}^T \mathbf{u}_{s',q}\), whereas Semi-Lagrangian employs a mapping form based on the ordinary advection equation \(\mathbf{u}_{s'\to r,q}^{A}=\mathbf{u}_{s',q}\).\color{black}

SMSP ensures robustness by addressing the two issues above: (1) The Neumann boundary can be set as \(\Lambda_{s',q}^r=\int_{s'}^r-\frac{1}{2}|\mathbf{u}_{\tau,q}|^2 d\tau\), namely, $\hat \lambda_q =-\frac{1}{2}|\mathbf{u}_{s',q}|^2\Delta t$, which is robustly calculable (because of only one timestep).
(2) The Poisson solve converges fast thanks to the small divergence on its right-hand side.
However, SMSP forgoes the advantages associated with employing a long-range flow map for vorticity conservation. Instead, it resolves to a reformulated Euler equation in covector form, accompanied by a one-step particle advection process. This approach results in the loss of the benefits inherent in utilizing a flow map to preserve vorticies.

\subsection{Long-Range Mapping, Short-Range Projection}\label{sec:first_solution}

A natural next step is to combine the merits of LMLP and SMSP. However, this is not mathematically intuitive. As shown in \autoref{eq:u_fm_particle}, the mapping and projection steps require the same time interval for their path integrals.  

To address this issue, we observe the following identity:
\begin{equation}
\boxed{\underbrace{\mathbf{u}_{s'\to r,q}^M}_{\text{SM}} = \underbrace{\mathbf{u}_{s\to r,q}^M}_{\text{LM}} - \nabla \Lambda_{s,q}^{s'}}.
\label{eq:relation_long_short}
\end{equation}

This identity can be simply proved as \(\mathbf{u}_{r,q} = \mathbf{u}_{s\to t,q}^M - \nabla \Lambda_{s,q}^r=\mathbf{u}_{s'\to t,q}^M - \nabla \Lambda_{s',q}^r\) and \( \Lambda_{s,q}^r= \Lambda_{s,q}^{s'}+\Lambda_{{s'},q}^r\). 
It establishes a connection between the long-range and short-range mappings, allowing us to express a short-range mapping by adding the gradient of pressure integral $\nabla \Lambda_{s,q}^{s'}$ to its long-range counterpart. This calculation is numerically robust, because for each particle, $\Lambda_{s,q}^{s'}$ is the path integral of the Lagrangian pressure on its trajectory. Substituting \autoref{eq:relation_long_short} into \autoref{eq:possion_2}, we obtain the following scheme:
\begin{enumerate}
    \item (\textbf{Long-Range Mapping}) Calculate mapped velocity by the long-range flow map: \({\mathbf{u}_{s\to r,q}^{M}} = {\mathcal{T}_{r,q}^s}^T \mathbf{u}_{s,q}\);
    \item Calculate mapped velocity by short-range flow map based on the long-range mapped velocity as: \({\mathbf{u}_{s'\to r,q}^{M}} = {\mathbf{u}_{s\to r,q}^{M}}-\nabla \Lambda_{s,q}^{s'}\)
    \item (\textbf{Short-Range Projection}) Solve the following Poisson equation to obtain $\hat \lambda_q$:
    \begin{equation}
    \begin{cases}
    \nabla\cdot\nabla \hat \lambda =\nabla\cdot {\mathbf{u}_{s'\to r}^{M}}, &  \mathbf{x}\in \Omega\\
     \mathbf{u}_r=\mathbf{u}_b, &  \mathbf{x}\in \partial_s \Omega\\
     p=0 \Rightarrow \hat \lambda = -\frac{1}{2}|\mathbf{u}_{r}^2|\Delta t, & \mathbf{x}\in \partial_f \Omega
    \end{cases}
    \label{eq:possion_3}
    \end{equation}
    where $\hat \lambda_q=\lambda_{r,q}\Delta t$ is the numerical calculation of one step integration $\Lambda_{s',q}^r.$
    \item Update the integral: $\Lambda_{s,q}^r=\Lambda_{s,q}^{s'}+\hat \lambda_q$ and project \(\mathbf{u}_{r,q} = {\mathbf{u}_{s\to r,q}^{M}} - \nabla \Lambda_{s,q}^r\) to ensure \(\mathbf{u}_{r}\) satisfies \(\nabla \cdot \mathbf{u}_r = 0\).
\end{enumerate}

We name it as \color{black}\textbf{L}ong-Range \textbf{M}apping \textbf{S}hort-Range \textbf{P}rojection\color{black} (\textbf{LMSP}). In LMSP, the $\Lambda_{s}^{s'}$ in Step (2) is calculated in the previous time step, and this value is updated in Step (4) by $\Lambda_{s,q}^r=\Lambda_{s,q}^{s'}+\hat \lambda_q$ to the value in the current time step. This accumulation might lead to a concern that accumulating pressure $\Lambda_s^r$ could potentially lead to the accumulation of numerical errors, thereby diminishing the long-range map's ability to preserve vorticity when calculating $\mathbf{u}_{s\to r,q}^M-\nabla \Lambda_s^{s'}$ in Step (2). 
We show this is not an issue.
Because $\nabla \Lambda_s^{s'}$ is a gradient field, it only affects the divergence component of $\mathbf{u}_{s\to r,q}^M$. Any numerical error accumulated in $\Lambda_s^{s'}$ goes directly into the rotational part $\mathbf{u}_{s\to r,q}^M$, which will be removed after projection. 

We further demonstrate this with the following proposition: 
\begin{proposition}\label{equivalence_poission}
    Poission \autoref{eq:possion_3} with initial guess $\hat \lambda =0 $ is equivalent to Poission \autoref{eq:possion_1} with initial guess $\Lambda_s^r=\Lambda_s^{s'}$
\end{proposition}
\paragraph{Proof:}  Substitute \autoref{eq:relation_long_short} into \autoref{eq:possion_3}, we obtain $\nabla \cdot \nabla \hat \lambda =\nabla \cdot [\nabla \mathbf{u}_{s\to r,q}^M-\nabla \Lambda_{s,q}^{s'}$], which is equivalent to $\nabla \cdot \nabla (\hat \lambda+ \Lambda_{s,q}^{s'}) =\nabla \cdot \nabla \mathbf{u}_{s\to r,q}^M$.  Use $\Lambda_s^r$ to substitute $\hat \lambda+ \Lambda_{s,q}^{s'}$, and we get \autoref{eq:possion_1}.  Also, for the initial guess, when $\hat \lambda = 0$, $\Lambda=\Lambda_{s,q}^{s'}$.

\section{Adapting to Classical Advection-Projection}
In the final movement, we further adapt the LMSP scheme to a classical advection-projection scheme by solving the Poisson equation with standard Neumann and Dirichlet boundary conditions, namely \(p=0\) on \(\partial_f\Omega\) and \(\bm u=\bm u_b\) on \(\partial_s\Omega\). This modification is motivated by the desire to circumvent any inaccuracies arising from the approximation of \(\mathcal{T}_s^r\) for a particle on the surface (due to the lack of sufficient neighboring particles to approximate the Jacobian), which could potentially lead to numerical instabilities during the simulation. We showed an ablation test for this issue in Fig.~\ref{fig:ablation_study_free_surface}.

We observe the following identity regarding a particle's velocity on a one-step flow map: 
\begin{equation}
    \boxed{\mathbf{u}_{s'\to r,q}^A=\mathbf{u}_{s'\to r,q}^M + \Delta t \color{black}\nabla (\frac{1}{2}|\mathbf{u}_{s',q}|^2) \color{black}}
    \label{eq:relation_short_common}
\end{equation}
where $\mathbf{u}_{s'\to r,q}^A$ represents the passively advected velocity on the particle's trajectory, namely, to move the particle to a new position and keep its velocity as it is, as seen in all traditional particle-based advection schemes. We show a brief proof in the \autoref{app:relation_short_common_proof}. 

Combining \autoref{eq:relation_long_short} and \autoref{eq:relation_short_common}, we obtain our final advection-projection scheme as:
\begin{enumerate}
    \item (\textbf{Long-Range Mapping}) For the interior particles, calculate the long-flow-map mapped velocity: \({\mathbf{u}_{s\to r,q}^{M}} = {\mathcal{T}_{r,q}^s}^T \mathbf{u}_{s,q}\).
    \item For an interior particle, calculate its velocity as the advected velocity expressed with long-flow-map mapped velocity (Eq.~\ref{eq:relation_short_common}): \({\mathbf{u}_{s'\to r,q}^{A}} = {\mathbf{u}_{s\to r}^{M}}-\nabla \Lambda_{s}^{s'}+\color{black}\nabla (\frac{1}{2}|\mathbf{u}_{s',q}|^2)\color{black} \Delta t\). 
    \item For a boundary particle (refer to \autoref{fig:particle_type}), calculate its velocity as the advected velocity: $\mathbf{u}_{s'\to r,q}^{A}=\mathbf{u}_{s',q}$. 
    \item (\textbf{Classical Projection}) Solve the classical Poisson equation to obtain $p_{q}$:
    \begin{equation}
    \begin{cases}
            \nabla \cdot \nabla p=\nabla\cdot \mathbf{u}_{s'\to r}^A, &  \mathbf{x}\in \Omega\\
     \mathbf{u}_r=\mathbf{u}_b, &  \mathbf{x}\in \partial_s \Omega\\
     p=0. & \mathbf{x}\in \partial_f \Omega
    \end{cases}
    \label{eq:possion_4}
    \end{equation}
    
    \item Update the integral: $\Lambda_{s,q}^r=\Lambda_{s,q}^{s'}+\Delta t(p_{q}-\frac{1}{2}|\mathbf{u}_{s',q}|^2)$ and do projection \(\mathbf{u}_{r,q} = {\mathbf{u}_{s\to r,q}^{M}} - \nabla \Lambda_{s,q}^r\) for interior part, and \(\mathbf{u}_{r,q} = {\mathbf{u}_{s'\to r,q}^{A}} - \nabla p_q\Delta t\) for the part near free surface.
\end{enumerate}

We name it as \textbf{L}ong-Range \textbf{M}apping \textbf{Classical} \textbf{P}rojection (\textbf{LMCP}). 
In this scheme, the calculation of the Poisson equation is performed for $u^A_{s'\to t}$ throughout the entire domain, ensuring the mathematical consistency of the final velocity by the covector flow map $u_{s\to r,q}^M$ and the advected velocity $u_{s\to r,q}^A$ after projection. At the same time, for interior particles, the advected velocity $u^A_{s'\to t}$ is calculated by \({\mathbf{u}_{s'\to r,q}^{A}} = {\mathbf{u}_{s\to r,q}^{M}}-\nabla \Lambda_{s,q}^{s'}+\color{black}\nabla (\frac{1}{2}|\mathbf{u}_{s',q}|^2)\color{black}\Delta t\) with long-range mapping for vorticity preserving. Again, because only the gradient $-\nabla \Lambda_{s}^{s'}+\color{black}\nabla (\frac{1}{2}|\mathbf{u}_{s'}|^2)\color{black} \Delta t$ is added to \({\mathbf{u}_{s\to r}^{M}}\), similar to \autoref{sec:first_solution}, the process of adding $-\nabla \Lambda_{s}^{s'}+\color{black}\nabla (\frac{1}{2}|\mathbf{u}_{s'}|^2)\color{black} \Delta t$ to \({\mathbf{u}_{s\to r}^{M}}\) to obtain \(\mathbf{u}_{s'\to r}^{A}\) keeps the long-range flow map's ability to preserve vorticity. 

\textbf{Summary.} We establish a long-range, Lagrangian flow map based on Equation~\ref{eq:u_fm_particle} and incorporate it in a classical projection step with zero Neumann boundary by leveraging Equation~\ref{eq:relation_long_short} and ~\ref{eq:relation_short_common}. 

\section{Time Integration}
\color{black}We adopt the LMCP scheme in our simulation \color{black}and summarize the time integration scheme of our approach in Algorithm~\autoref{alg:final_algorithm}. We provide further implementation details in the \autoref{app:details}. 

\begin{algorithm}[h]
\caption{Lagrangian Covector Fluid}
\label{alg:final_algorithm}
\begin{flushleft}
        \textbf{Input:} Initial velocity $\mathbf{u}_s$; Initial particle positions $\mathbf{x}_s$; \color{black}Re-initialization decision strategy $\mathcal{R}$\color{black};  Near-surface judgment$\mathcal{J}$
\end{flushleft}
\begin{algorithmic}[1]
\State set time $s \gets 0, s'\gets 0$; integration $\Lambda_{s,i}^s\gets 0$
\For{\textbf{each} time step and at time $r$}
     \If{re-initialization decision strategy $\mathcal{R}$ satisfy}
        \State set $s \gets r,s' \gets r$
        \State set $\mathbf{u}_{s,i} \gets \mathbf{u}_{i}$; 
 $\mathbf{x}_{s,i} \gets \mathbf{x}_{i}$; $\Lambda_{s,i}^s\gets 0$ 
     \EndIf{}
    \For{\textbf{all} particle $i$}
        \State advect particle position: $\mathbf{x}_i \gets \mathbf{x}_i+ \mathbf{v}_i \Delta t$
    \EndFor{}
    \For{\textbf{all} particle $i$}
    \If{$\mathcal{J}(i)$ is True}
        \State compute ${\mathbf{u}_{s'\to r,i}^A} \gets u_{s',i}$
    \Else
    \State compute $\mathcal{T}_{r,i}^s\gets\frac{\partial \mathbf{x}_s}{\partial \mathbf{x}_r}|_{\mathbf{x}_r=\mathbf{x}_i}$
    \State compute ${\mathbf{u}_{s\to r,i}^M}\gets \mathcal{T}_{r,i}^s \mathbf{u}_{s,i}$
    \State compute ${\mathbf{u}_{s'\to r,i}^A}\gets{\mathbf{u}_{s\to r,i}^M}-\color{black}\nabla\color{black} \Lambda_{s,i}^{s'}+\color{black}\nabla (\frac{1}{2}|\mathbf{u}_{s',i}|^2)\color{black}\Delta t$ 
    \EndIf{}
 
    \EndFor{}    
    \State solve possion \autoref{eq:possion_4} and get $p_{r,i}$
    \State update $\Lambda_{s,i}^r\gets \Lambda_{s,i}^{s'}+(p_{r,i}-\frac{1}{2}|\mathbf{u}_{r,i}|^2)\Delta t$
    \For{\textbf{all} particle $i$}
        \State do velocity projection: $\mathbf{u}_{r,i}=\mathbf{u}_{s'\to r,i}^A-\nabla p_{r,i}$
    \EndFor{}
    \State set last time: $s'\gets r$
\EndFor{}
\end{algorithmic}
\end{algorithm}

\section{Numerical Implementation}

\paragraph{Voronoi-based Discretization}

\begin{wrapfigure}[11]{r}{0.2\textwidth}
    \centering
    \vspace{-0.5cm}
    \includegraphics[width=0.2\textwidth]{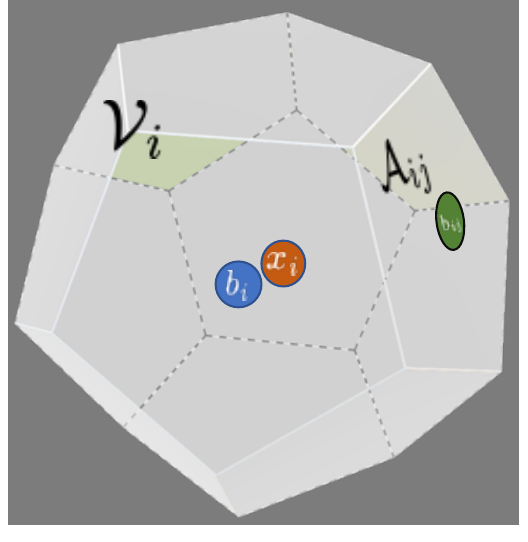}
    \vspace{-0.7cm}\caption{Voronoi discretization}\label{fig:voronoi}
\end{wrapfigure}

We implemented a Voronoi-based particle method for numerical implementation. In each time step, we generate Voronoi diagrams for all moving particles, with each particle corresponding to a Voronoi cell.  We represent the $i$-th particle as $q_i$, with position $\mathbf{x}_i$.  As shown in Fig.~\ref{fig:voronoi}, the cell corresponding to $q_i$ is denoted $\mathcal{V}_i$, with a volume $V_i$ and centroid $\mathbf{b}_i$. The adjacent facet between two neighboring cells \color{black}$\mathcal{V}_i$ and $\mathcal{V}_j$ \color{black}is denoted as $\mathcal{A}_{ij}$. The facet $\mathcal{A}_{ij}$ has the area $A_{ij}$ and the centroid $\mathbf{b}_{ij}$. The distances from $\mathbf{x}_i$ and $\mathbf{x}_j$ to the facet $\mathcal{A}_{ij}$ are respectively denoted as $d_{ij}$ and $d_{ji}$, while the distance between $\mathbf{x}_i$ and $\mathbf{x}_j$ is denoted as $l_{ij}$. According to the properties of Voronoi diagrams, $\mathcal{A}_{ij}$ bisects the line connecting $\mathbf{x}_i$ and $\mathbf{x}_j$ perpendicularly, so $d_{ij} = d_{ji}$ and $d_{ij} + d_{ji} = l_{ij}$ holds.

For each particle $q_i$, its associated Voronoi cell $\mathcal{V}_i$ is used to define \color{black}a matrix-form discrete gradient operator $G$ and divergence operator $D$ to calculate the gradient $Gp$ of a scalar quantity $q$ and the divergence $D\mathbf{v}$ of a vector quantity $\mathbf{v}$   \color{black}carried by particles. The computation of these operators relies on the calculation of the rate of change of the volume $\nabla_{\mathbf{x}_i} V_j$
induced by particle positions. We use the formula $    \nabla_{\mathbf{x}_j} V_i=\frac{A_{ij}}{l_{ij}}(\mathbf{x}_j-\mathbf{b}_{ij})$ and $\nabla_{\mathbf{x}_i} V_i=-\sum_{j\in\mathcal{N}_i} \nabla_{\mathbf{x}_i} V_j$ for calculating $\nabla_{\mathbf{x}_i} V_j$ as given in \cite{de2015power}.

According to \cite{de2015power,duque2023unified}, the matrix-form divergence operator and gradient operator can be defined directly from $\nabla_{\mathbf{x}_i} V_j$ as $D_{ij}=(\nabla_{\mathbf{x}_j} V_i)^T$, $D_{ii}=(\nabla_{\mathbf{x}_i} V_i)^T$ and $G=-D^T$.  
The divergence of $\mathbf{v}$ and the gradient of $p$ can be computed using the matrix-form divergence and gradient operators defined as follows:
\begin{equation}
    \begin{aligned}
        [D\mathbf{v}]_i&= \sum_{j\in N_i} \frac{A_{ij}}{l_{ij}}[(\mathbf{b}_{ij}-\mathbf{x}_i)\cdot \mathbf{v}_i+(\mathbf{x}_j-\mathbf{b}_{ij})\cdot \mathbf{v}_j],\\
        [Gp]_i&= \sum_{j\in N_i} \frac{A_{ij}}{l_{ij}}(\mathbf{x}_i-\mathbf{b}_{ij})(p_i-p_j),
    \end{aligned}
\end{equation}
where $N_i$ represents the set composed of the cells adjacent to cell $i$.

When solving the Poisson equation for velocity projection, the Laplacian operator $L$ is defined as $L = DV^{-1}G$, where $V$ is a diagonal matrix composed of all cell volumes $V_i$. Because the differential operator matrices satisfy $G = -D^T$, the operator $-L$ is symmetric and positive semi-definite, allowing the discretized Poisson equation $-L p=-D \mathbf{u}^*$ to be solved using the Conjugate Gradient method, where $\mathbf{u}^*$ represents the velocity before the divergence projection. 

\begin{wrapfigure}[10]{R}{0.2\textwidth}
    \centering
    \vspace{-0.5cm}
    \hspace{-1.5cm}
    \includegraphics[width=0.22\textwidth]{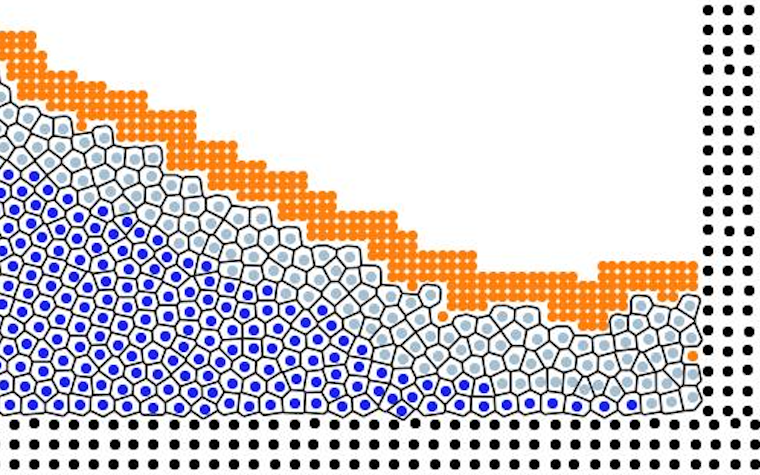}
    \vspace{-0.4cm}
    \hspace{-1.5cm}
    \caption{Particles represent either interior (blue) or boundary (gray-blue) fluid, boundaries (black), or air (orange).\label{fig:particle_type}}
\end{wrapfigure}



\paragraph{Boundary} Similar to \cite{de2015power}, we represent the solid boundary using solid particles and sample air particles near the free surface, utilizing the ghost particle method from \cite{schechter2012ghost} (see Fig.~\ref{fig:particle_type}). Solid particles and air particles are used for clipping the Voronoi diagrams calculated for fluid particles and setting boundary conditions, as proposed in \cite{de2015power}.

\color{black}\paragraph{Others}\color{black} For gravity, we accumulate the integration of gravity $G_s^r=\int_s^r g d\tau$ along the particle trajectory and add it to the mapped velocity before projection:\color{black}${\mathbf{u}^M_{s\to r}} \gets {\mathbf{u}^M_{s\to r}}+G_s^{r}$ \color{black}for interior part and add gravity directly to advected velocity \color{black}${\mathbf{u}^A_{s'\to r}} \gets {\mathbf{u}^A_{s'\to r}}+g\Delta t$ \color{black} near the free surface.  To make particle distribution more uniform, like in \cite{de2015power}, at the end of each time step, we put particles to the centroid of their corresponding Voronoi cell $\mathbf{x}_i\gets \mathbf{b}_i$.

\section{Results and Discussion}\label{sec:validation}

\paragraph{Validation}
We demonstrate the effectiveness of our method by performing four benchmark experiments against the power particle method (PPM)~\cite{de2015power}\footnote{For simplicity, we implement a Voronoi diagram instead of the power diagram without affecting the ability to preserve vortex for both our methods and PPM.}, which shares implementation-wise similarities given its geometric data structure. We observe slower energy dissipation rates, less vorticity noise, and better preservation of vortical structures. We color each 2D particle blue (lower/negative values) through gray to red (higher/positive values), based on a linear curve corresponding to its vorticity magnitude.

\textbf{(1) Leapfrog.} 
We initialize two negative and two positive vortex rings, letting the four vortices move forward by the velocity field generated by their influence on each other. The rate of energy dissipation directly relates to the number of cycles they move forward before merging. As illustrated in Fig.~\ref{fig:leapfrog_2d}, our method shows a better preservation of the vortical structures than PPM; vortices simulated with our method stay separated at the point where the vortices in the PPM simulation already merged.
\textbf{(2) Taylor Vortices.} 
We benchmark our method by simulating two Taylor vortices placed $0.815$ apart from each other (Fig.~\ref{fig:2d_taylor}). Their velocity field is given by $\omega(\color{black}\mathbf{x}\color{black}) = U/a (2 - r^2 / a^2)exp((1 - r^2/a^2)/2)$, where we use $U = 1$ and $a = 0.3$, and $r$ denotes the distance from \color{black}$\mathbf{x}$ \color{black}to the vortex center.Using our method, we observe the vortices staying separate, whereas using PPM, they merge at the center.
\textbf{(3) Taylor-Green Vortices.}
Fig.~\ref{fig:taylor_green_2d} shows a simulation started with a symmetric divergence-free velocity field. The fluid is expected to maintain symmetry along the two axes in 2D while rotating. We observe our method producing less noise in terms of vorticity magnitude carried by the particles as compared to PPM (Fig.~\ref{fig:taylor_green_2d}a-b), with our method presenting a far better energy dissipation curve (Fig.~\ref{fig:taylor_green_2d}c).
\textbf{(4) 3D Dam Break.}
In Fig.~\ref{fig:dam_break}, we validate our method with the classical 3D benchmark case for verification of solid boundary and free surface flow handling.



\paragraph{Ablation Study}
We illustrate (1) the robustness of our LMCP scheme in handling free surface boundary and (2) the subtraction of accumulated pressure gradient significantly accelerating the convergence rate. 
Fig.~\ref{fig:ablation_study_free_surface} illustrates a robust interface achieved using our scheme, in contrast to instabilities that occur without it.
Fig.~\ref{fig:ablation_convergence} shows that the subtraction of the accumulated pressure gradient can accelerate the convergence of the Poisson equation solver.

\paragraph{Examples}We show additional 2D and 3D examples to demonstrate the robustness and correctness of our method. We use Taichi \cite{hu2019taichi} for our implementation, and experiments are run on Tesla V100 GPUs. We use at most $200\,000$ particles in all experiments to represent fluid, air, and solids. Voronoi diagrams are created using Scipy \cite{scipy2020} and Qhull \cite{barber1996quickhull}.
\textbf{K\'arm\'an Vortex Street.}
Fig.~\ref{fig:karman} shows alternating vortices forming downstream from a blunt object caused by the unsteady separation of the fluid.
\textbf{2D Moving and Rotating Board}
Fig.~\ref{fig:board_2d} shows an object exhibiting flapping motion by traversing the rectangular domain from left to right, and back while generating vortices in its wake.
\textbf{3D Single \& Double Sink}
\duowen{In these examples, we illustrate our method's ability for accurate vorticity perservation combining with free-surface treatment. For single vortex example, we place an initial vorticity field at the center of the tank. A hole is opened as the sink for the tank and water drains out through the hole. Similar settings are adopted for two sink but the sinks have opposite direction of rotation in order to create interesting surface motion. We can observe spiral patterns on the surface in both examples. Results are shown in Fig.~\ref{fig:one_sink_both},  Fig.~\ref{fig:two_sink_surface} and Fig.~\ref{fig:two_sink_particles}}
\textbf{3D Rotating Board}
\duowen{As illustrated in Fig.~\ref{fig:board_rotate}, a board is placed at the center of the scene and set to rotate at a constant speed. We show our method can handle drastic free-surface change and we deal moving solid boundaries in a robust and effective way. Splashes and detailed water surfaces can be observed.}
\textbf{3D Wave Generator.}
\duowen{
In this example, we demonstrate the scenario of waves crashing against several pillars. We observe the dynamic water flow behind the pillars and the interaction between the waves. We show particle view for this example in Fig.~\ref{fig:wave-columns}.} 

\color{black}\paragraph{CFL} In all of our examples, we set the CFL number to $1$ due to constraints imposed by the Voronoi particles. Larger CFL numbers could lead to drastic changes in the neighbors of particles, potentially causing stability issues. Grid-based methods \cite{nabizadeh2022covector} do not encounter these issues.
\color{black}

\begin{table}[ht]
\centering
\caption{\color{black}Performance timing. We measure the time each substep takes, and also how much of this was taken up by constructing the Voronoi diagram. Although our simulation method runs on the GPU, the Voronoi diagram calculation takes place on the CPU.
\iffalse We measure the average time cost per substep of our implementation. We calcualte the substeps needed for a given time step based on the CFL condition. The timings are obtained with a system using a Tesla V100 GPU (32Gb) and Dual Intel Xeon Gold 6226 CPUs (2.7 GHz).\fi\color{black}}
\vspace{-0.2cm}
\begin{tabular}{|l|c|c|}
\hline
\multicolumn{3}{|c|}{\textbf{Average Time Cost per Substep}} 
\\ \hline
\multicolumn{1}{|l|}{\textbf{Scene name}} & \textbf{Particles} & \textbf{Total (Voronoi)}
\\ \hline
2D Leapfrog                     & 480k & 12.5s (8.8s) \\
2D Taylor Vortices              & 360k & 9.8s  (6.4s) \\
2D Taylor-Green Vortices        & 156k & 5.96s (2.7s) \\
3D Dam Break                 & 211k & 20.9s (18.12s) \\
2D Kármán Vortex Street         & 381k & 9.56s (6.74s) \\
2D Moving and Rotating Board & 150k & 5.92s (2.66s) \\
3D Sink                      & 180k & 29s (25.7s) \\
3D Rotating Board            & 150k & 5.9s (2.6s) \\
3D Wave Generator            & 403k & 35s (32.84s) \\
\hline
\end{tabular}

\label{table-benchmark}
\vspace{-0.5cm}
\end{table}

\section{Limitations and Future Work}
In summary, this paper presents a novel Lagrangian approach to establishing covector flow maps under complex boundary conditions. 
The developed decoupling mechanism, rooted in flow-map theory, effectively combines long-range flow maps with short-range (and classical) projections, ensuring robust handling of free boundaries. 

A significant limitation of our approach lies in its exclusive treatment of inviscid flows. Addressing viscous flows, as well as other interfacial phenomena, represents a promising direction for future research. Currently, the speed of our fluid simulation code is constrained by the single-threaded Qhull algorithm used for generating Voronoi cells in each frame. We plan to investigate more efficient schemes for solving incompressibility on particles. In our future work, we aim to delve further into flow-map theories within a weakly compressible framework, enhancing meshfree Lagrangian methods such as SPH. Additionally, we are interested in applying our decoupled mapping-projection scheme to other free-surface problems, including levelset-based and particle-grid methods.

 \begin{acks}
We acknowledge NSF IIS \#2313075, ECCS \#2318814, CAREER \#2420319, IIS \#2106733, OISE \#2153560, and CNS \#1919647 for funding support. We credit the Houdini education license for the production of the video animations.
\end{acks}

\appendix

\section{Proof of Eq.~\ref{eq:relation_short_common}}
\label{app:relation_short_common_proof}
\paragraph{Proof:}  Consider one step advection from time $s'$ to $r$ and we have $\mathbf{x}_{r,q}=\mathbf{x}_{s',q}+\Delta t \mathbf{u}_{s',q}$.  Thus the Jacobian $\mathcal{T}_r^{s'}=\frac{\partial \mathbf{x}_{s',q}}{\partial \mathbf{x}_{r,q}}$ can be calculated as $\mathcal{T}_r^{s'}=\frac{\partial \mathbf{x}_{s',q}}{\partial \mathbf{x}_{r,q}}=\frac{\partial (\mathbf{x}_{r,q}-\Delta t \mathbf{u}_{s',q})}{\partial \mathbf{x}_{r,q}}=I-\Delta t \nabla \mathbf{u}_{s',q}$, where $I$ denotes identity matrix.  Thus $\mathbf{u}_{s'\to r,q}^M={\mathcal{T}_{s',q}^{s'}}^T \mathbf{u}_{s',q}=\mathbf{u}_{s',q}-\Delta t \nabla \mathbf{u}_{s',q}^T\cdot \mathbf{u}_{s',q}$.  In the particle method, $\mathbf{u}_{s'\to r,q}^A=\mathbf{u}_{s',q}$. Due to $\nabla \mathbf{u}^T\cdot \mathbf{u}=\color{black}\nabla (\frac{1}{2}|\mathbf{u}|^2)\color{black}$, we have $\mathbf{u}_{s'\to r,q}^A=\mathbf{u}_{s'\to r,q}^M+\color{black}\nabla (\frac{1}{2}|\mathbf{u_{s',q}}|^2)\color{black} \Delta t$. \hfill $\square$
\section{Implementation Details of Algo.~\ref{alg:final_algorithm}}
\label{app:details}

\textbf{Reinitialization.} \color{black}We employ a simple reinitialization decision strategy $\mathcal{R}$ triggered every $n$ substeps ($n=20$ in our implementation). \color{black}

\noindent\textbf{Boundary particle checking.} We employ the following strategy to obtain the boundary particle set $\mathcal{J}$: At the initial time $s$, set the flag $\mathscr{j}_i$ to False; at each step, check if the particle is in the $k$ layers of particles near the free surface at current time $r$. If it is, we update the flag $\mathscr{j}_i$ to True; $J(i)$ returns the value of $\mathscr{j}_i$.(\color{black}$k=3$ in our implementation\color{black}).


\bibliographystyle{ACM-Reference-Format}
\bibliography{refs_ML_sim.bib, refs_INR.bib, refs_flow_map.bib, refs_simulation.bib}


\begin{thebibliography}{34}


\ifx \showCODEN    \undefined \def \showCODEN     #1{\unskip}     \fi
\ifx \showDOI      \undefined \def \showDOI       #1{#1}\fi
\ifx \showISBNx    \undefined \def \showISBNx     #1{\unskip}     \fi
\ifx \showISBNxiii \undefined \def \showISBNxiii  #1{\unskip}     \fi
\ifx \showISSN     \undefined \def \showISSN      #1{\unskip}     \fi
\ifx \showLCCN     \undefined \def \showLCCN      #1{\unskip}     \fi
\ifx \shownote     \undefined \def \shownote      #1{#1}          \fi
\ifx \showarticletitle \undefined \def \showarticletitle #1{#1}   \fi
\ifx \showURL      \undefined \def \showURL       {\relax}        \fi
\providecommand\bibfield[2]{#2}
\providecommand\bibinfo[2]{#2}
\providecommand\natexlab[1]{#1}
\providecommand\showeprint[2][]{arXiv:#2}

\bibitem[\protect\citeauthoryear{Barber, Dobkin, and Huhdanpaa}{Barber et~al\mbox{.}}{1996}]%
        {barber1996quickhull}
\bibfield{author}{\bibinfo{person}{C.B. Barber}, \bibinfo{person}{D.P. Dobkin}, {and} \bibinfo{person}{H.T. Huhdanpaa}.} \bibinfo{year}{1996}\natexlab{}.
\newblock \showarticletitle{The Quickhull algorithm for convex hulls}.
\newblock \bibinfo{journal}{\emph{ACM Transactions on Mathematical Software (TOMS)}} \bibinfo{volume}{22}, \bibinfo{number}{4} (\bibinfo{date}{Dec} \bibinfo{year}{1996}), \bibinfo{pages}{469--483}.
\newblock
\urldef\tempurl%
\url{http://www.qhull.org}
\showURL{%
\tempurl}


\bibitem[\protect\citeauthoryear{Buttke}{Buttke}{1992}]%
        {buttke1992lagrangian}
\bibfield{author}{\bibinfo{person}{TF Buttke}.} \bibinfo{year}{1992}\natexlab{}.
\newblock \showarticletitle{Lagrangian numerical methods which preserve the Hamiltonian structure of incompressible fluid flow}.
\newblock  (\bibinfo{year}{1992}).
\newblock


\bibitem[\protect\citeauthoryear{Buttke}{Buttke}{1993}]%
        {buttke1993velicity}
\bibfield{author}{\bibinfo{person}{Tomas~F Buttke}.} \bibinfo{year}{1993}\natexlab{}.
\newblock \showarticletitle{Velicity methods: Lagrangian numerical methods which preserve the Hamiltonian structure of incompressible fluid flow}.
\newblock In \bibinfo{booktitle}{\emph{Vortex flows and related numerical methods}}. \bibinfo{publisher}{Springer}, \bibinfo{pages}{39--57}.
\newblock


\bibitem[\protect\citeauthoryear{Buttke and Chorin}{Buttke and Chorin}{1993}]%
        {buttke1993turbulence}
\bibfield{author}{\bibinfo{person}{Thomas~F Buttke} {and} \bibinfo{person}{Alexandre~J Chorin}.} \bibinfo{year}{1993}\natexlab{}.
\newblock \showarticletitle{Turbulence calculations in magnetization variables}.
\newblock \bibinfo{journal}{\emph{Applied numerical mathematics}} \bibinfo{volume}{12}, \bibinfo{number}{1-3} (\bibinfo{year}{1993}), \bibinfo{pages}{47--54}.
\newblock


\bibitem[\protect\citeauthoryear{Cortez}{Cortez}{1996}]%
        {cortez1996impulse}
\bibfield{author}{\bibinfo{person}{Ricardo Cortez}.} \bibinfo{year}{1996}\natexlab{}.
\newblock \showarticletitle{An impulse-based approximation of fluid motion due to boundary forces}.
\newblock \bibinfo{journal}{\emph{J. Comput. Phys.}} \bibinfo{volume}{123}, \bibinfo{number}{2} (\bibinfo{year}{1996}), \bibinfo{pages}{341--353}.
\newblock


\bibitem[\protect\citeauthoryear{Crane, de~Goes, Desbrun, and Schr\"{o}der}{Crane et~al\mbox{.}}{2013}]%
        {Crane2013Exteriorcalculus}
\bibfield{author}{\bibinfo{person}{Keenan Crane}, \bibinfo{person}{Fernando de Goes}, \bibinfo{person}{Mathieu Desbrun}, {and} \bibinfo{person}{Peter Schr\"{o}der}.} \bibinfo{year}{2013}\natexlab{}.
\newblock \showarticletitle{Digital geometry processing with discrete exterior calculus}. In \bibinfo{booktitle}{\emph{ACM SIGGRAPH 2013 Courses}} (Anaheim, California) \emph{(\bibinfo{series}{SIGGRAPH '13})}. \bibinfo{publisher}{Association for Computing Machinery}, \bibinfo{address}{New York, NY, USA}, Article \bibinfo{articleno}{7}, \bibinfo{numpages}{126}~pages.
\newblock
\showISBNx{9781450323390}
\urldef\tempurl%
\url{https://doi.org/10.1145/2504435.2504442}
\showDOI{\tempurl}


\bibitem[\protect\citeauthoryear{De~Goes, Wallez, Huang, Pavlov, and Desbrun}{De~Goes et~al\mbox{.}}{2015}]%
        {de2015power}
\bibfield{author}{\bibinfo{person}{Fernando De~Goes}, \bibinfo{person}{Corentin Wallez}, \bibinfo{person}{Jin Huang}, \bibinfo{person}{Dmitry Pavlov}, {and} \bibinfo{person}{Mathieu Desbrun}.} \bibinfo{year}{2015}\natexlab{}.
\newblock \showarticletitle{Power particles: an incompressible fluid solver based on power diagrams.}
\newblock \bibinfo{journal}{\emph{ACM Trans. Graph.}} \bibinfo{volume}{34}, \bibinfo{number}{4} (\bibinfo{year}{2015}), \bibinfo{pages}{50--1}.
\newblock


\bibitem[\protect\citeauthoryear{Deng, Yu, Zhang, Wu, and Zhu}{Deng et~al\mbox{.}}{2023}]%
        {deng2023fluid}
\bibfield{author}{\bibinfo{person}{Yitong Deng}, \bibinfo{person}{Hong-Xing Yu}, \bibinfo{person}{Diyang Zhang}, \bibinfo{person}{Jiajun Wu}, {and} \bibinfo{person}{Bo Zhu}.} \bibinfo{year}{2023}\natexlab{}.
\newblock \showarticletitle{Fluid Simulation on Neural Flow Maps}.
\newblock \bibinfo{journal}{\emph{ACM Transactions on Graphics (TOG)}} \bibinfo{volume}{42}, \bibinfo{number}{6} (\bibinfo{year}{2023}), \bibinfo{pages}{1--21}.
\newblock


\bibitem[\protect\citeauthoryear{Duque}{Duque}{2023}]%
        {duque2023unified}
\bibfield{author}{\bibinfo{person}{Daniel Duque}.} \bibinfo{year}{2023}\natexlab{}.
\newblock \showarticletitle{A unified derivation of Voronoi, power, and finite-element Lagrangian computational fluid dynamics}.
\newblock \bibinfo{journal}{\emph{European Journal of Mechanics-B/Fluids}}  \bibinfo{volume}{98} (\bibinfo{year}{2023}), \bibinfo{pages}{268--278}.
\newblock


\bibitem[\protect\citeauthoryear{Feng, Liu, Xiong, Yang, Zhang, and Zhu}{Feng et~al\mbox{.}}{2022}]%
        {feng2022impulse}
\bibfield{author}{\bibinfo{person}{Fan Feng}, \bibinfo{person}{Jinyuan Liu}, \bibinfo{person}{Shiying Xiong}, \bibinfo{person}{Shuqi Yang}, \bibinfo{person}{Yaorui Zhang}, {and} \bibinfo{person}{Bo Zhu}.} \bibinfo{year}{2022}\natexlab{}.
\newblock \showarticletitle{Impulse fluid simulation}.
\newblock \bibinfo{journal}{\emph{IEEE Transactions on Visualization and Computer Graphics}} (\bibinfo{year}{2022}).
\newblock


\bibitem[\protect\citeauthoryear{Hachisuka}{Hachisuka}{2005}]%
        {hachisuka2005combined}
\bibfield{author}{\bibinfo{person}{Toshiya Hachisuka}.} \bibinfo{year}{2005}\natexlab{}.
\newblock \showarticletitle{Combined Lagrangian-Eulerian approach for accurate advection}.
\newblock In \bibinfo{booktitle}{\emph{ACM SIGGRAPH 2005 Posters}}. \bibinfo{pages}{114--es}.
\newblock


\bibitem[\protect\citeauthoryear{Hu, Li, Anderson, Ragan-Kelley, and Durand}{Hu et~al\mbox{.}}{2019}]%
        {hu2019taichi}
\bibfield{author}{\bibinfo{person}{Yuanming Hu}, \bibinfo{person}{Tzu-Mao Li}, \bibinfo{person}{Luke Anderson}, \bibinfo{person}{Jonathan Ragan-Kelley}, {and} \bibinfo{person}{Fr{\'e}do Durand}.} \bibinfo{year}{2019}\natexlab{}.
\newblock \showarticletitle{Taichi: a language for high-performance computation on spatially sparse data structures}.
\newblock \bibinfo{journal}{\emph{ACM Transactions on Graphics (TOG)}} \bibinfo{volume}{38}, \bibinfo{number}{6} (\bibinfo{year}{2019}), \bibinfo{pages}{201}.
\newblock


\bibitem[\protect\citeauthoryear{L{\'e}vy}{L{\'e}vy}{2022}]%
        {levy2022partial}
\bibfield{author}{\bibinfo{person}{Bruno L{\'e}vy}.} \bibinfo{year}{2022}\natexlab{}.
\newblock \showarticletitle{Partial optimal transport for a constant-volume Lagrangian mesh with free boundaries}.
\newblock \bibinfo{journal}{\emph{J. Comput. Phys.}}  \bibinfo{volume}{451} (\bibinfo{year}{2022}), \bibinfo{pages}{110838}.
\newblock


\bibitem[\protect\citeauthoryear{Mercier, Yin, and Nave}{Mercier et~al\mbox{.}}{2020}]%
        {mercier2020characteristic}
\bibfield{author}{\bibinfo{person}{Olivier Mercier}, \bibinfo{person}{Xi-Yuan Yin}, {and} \bibinfo{person}{Jean-Christophe Nave}.} \bibinfo{year}{2020}\natexlab{}.
\newblock \showarticletitle{The characteristic mapping method for the linear advection of arbitrary sets}.
\newblock \bibinfo{journal}{\emph{SIAM Journal on Scientific Computing}} \bibinfo{volume}{42}, \bibinfo{number}{3} (\bibinfo{year}{2020}), \bibinfo{pages}{A1663--A1685}.
\newblock


\bibitem[\protect\citeauthoryear{Nabizadeh, Wang, Ramamoorthi, and Chern}{Nabizadeh et~al\mbox{.}}{2022}]%
        {nabizadeh2022covector}
\bibfield{author}{\bibinfo{person}{Mohammad~Sina Nabizadeh}, \bibinfo{person}{Stephanie Wang}, \bibinfo{person}{Ravi Ramamoorthi}, {and} \bibinfo{person}{Albert Chern}.} \bibinfo{year}{2022}\natexlab{}.
\newblock \showarticletitle{Covector fluids}.
\newblock \bibinfo{journal}{\emph{ACM Transactions on Graphics (TOG)}} \bibinfo{volume}{41}, \bibinfo{number}{4} (\bibinfo{year}{2022}), \bibinfo{pages}{1--16}.
\newblock


\bibitem[\protect\citeauthoryear{Oseledets}{Oseledets}{1989}]%
        {oseledets1989new}
\bibfield{author}{\bibinfo{person}{Valery~Iustinovich Oseledets}.} \bibinfo{year}{1989}\natexlab{}.
\newblock \showarticletitle{On a new way of writing the Navier-Stokes equation. The Hamiltonian formalism}.
\newblock \bibinfo{journal}{\emph{Russ. Math. Surveys}}  \bibinfo{volume}{44} (\bibinfo{year}{1989}), \bibinfo{pages}{210--211}.
\newblock


\bibitem[\protect\citeauthoryear{Qu, Zhang, Gao, Jiang, and Chen}{Qu et~al\mbox{.}}{2019}]%
        {qu2019efficient}
\bibfield{author}{\bibinfo{person}{Ziyin Qu}, \bibinfo{person}{Xinxin Zhang}, \bibinfo{person}{Ming Gao}, \bibinfo{person}{Chenfanfu Jiang}, {and} \bibinfo{person}{Baoquan Chen}.} \bibinfo{year}{2019}\natexlab{}.
\newblock \showarticletitle{Efficient and conservative fluids using bidirectional mapping}.
\newblock \bibinfo{journal}{\emph{ACM Transactions on Graphics (TOG)}} \bibinfo{volume}{38}, \bibinfo{number}{4} (\bibinfo{year}{2019}), \bibinfo{pages}{1--12}.
\newblock


\bibitem[\protect\citeauthoryear{Roberts}{Roberts}{1972}]%
        {roberts1972hamiltonian}
\bibfield{author}{\bibinfo{person}{PH Roberts}.} \bibinfo{year}{1972}\natexlab{}.
\newblock \showarticletitle{A Hamiltonian theory for weakly interacting vortices}.
\newblock \bibinfo{journal}{\emph{Mathematika}} \bibinfo{volume}{19}, \bibinfo{number}{2} (\bibinfo{year}{1972}), \bibinfo{pages}{169--179}.
\newblock


\bibitem[\protect\citeauthoryear{Sato, Batty, Igarashi, and Ando}{Sato et~al\mbox{.}}{2018}]%
        {sato2018spatially}
\bibfield{author}{\bibinfo{person}{Takahiro Sato}, \bibinfo{person}{Christopher Batty}, \bibinfo{person}{Takeo Igarashi}, {and} \bibinfo{person}{Ryoichi Ando}.} \bibinfo{year}{2018}\natexlab{}.
\newblock \showarticletitle{Spatially adaptive long-term semi-Lagrangian method for accurate velocity advection}.
\newblock \bibinfo{journal}{\emph{Computational Visual Media}} \bibinfo{volume}{4}, \bibinfo{number}{3} (\bibinfo{year}{2018}), \bibinfo{pages}{6}.
\newblock


\bibitem[\protect\citeauthoryear{Sato, Igarashi, Batty, and Ando}{Sato et~al\mbox{.}}{2017}]%
        {sato2017long}
\bibfield{author}{\bibinfo{person}{Takahiro Sato}, \bibinfo{person}{Takeo Igarashi}, \bibinfo{person}{Christopher Batty}, {and} \bibinfo{person}{Ryoichi Ando}.} \bibinfo{year}{2017}\natexlab{}.
\newblock \showarticletitle{A long-term semi-lagrangian method for accurate velocity advection}.
\newblock In \bibinfo{booktitle}{\emph{SIGGRAPH Asia 2017 Technical Briefs}}. \bibinfo{pages}{1--4}.
\newblock


\bibitem[\protect\citeauthoryear{Saye}{Saye}{2016}]%
        {saye2016interfacial}
\bibfield{author}{\bibinfo{person}{Robert Saye}.} \bibinfo{year}{2016}\natexlab{}.
\newblock \showarticletitle{Interfacial gauge methods for incompressible fluid dynamics}.
\newblock \bibinfo{journal}{\emph{Science advances}} \bibinfo{volume}{2}, \bibinfo{number}{6} (\bibinfo{year}{2016}), \bibinfo{pages}{e1501869}.
\newblock


\bibitem[\protect\citeauthoryear{Saye}{Saye}{2017}]%
        {saye2017implicit}
\bibfield{author}{\bibinfo{person}{Robert Saye}.} \bibinfo{year}{2017}\natexlab{}.
\newblock \showarticletitle{Implicit mesh discontinuous Galerkin methods and interfacial gauge methods for high-order accurate interface dynamics, with applications to surface tension dynamics, rigid body fluid--structure interaction, and free surface flow: Part I}.
\newblock \bibinfo{journal}{\emph{J. Comput. Phys.}}  \bibinfo{volume}{344} (\bibinfo{year}{2017}), \bibinfo{pages}{647--682}.
\newblock


\bibitem[\protect\citeauthoryear{Schechter and Bridson}{Schechter and Bridson}{2012}]%
        {schechter2012ghost}
\bibfield{author}{\bibinfo{person}{Hagit Schechter} {and} \bibinfo{person}{Robert Bridson}.} \bibinfo{year}{2012}\natexlab{}.
\newblock \showarticletitle{Ghost SPH for animating water}.
\newblock \bibinfo{journal}{\emph{ACM Transactions on Graphics (TOG)}} \bibinfo{volume}{31}, \bibinfo{number}{4} (\bibinfo{year}{2012}), \bibinfo{pages}{1--8}.
\newblock


\bibitem[\protect\citeauthoryear{Summers}{Summers}{2000}]%
        {summers2000representation}
\bibfield{author}{\bibinfo{person}{DM Summers}.} \bibinfo{year}{2000}\natexlab{}.
\newblock \showarticletitle{A representation of bounded viscous flow based on Hodge decomposition of wall impulse}.
\newblock \bibinfo{journal}{\emph{J. Comput. Phys.}} \bibinfo{volume}{158}, \bibinfo{number}{1} (\bibinfo{year}{2000}), \bibinfo{pages}{28--50}.
\newblock


\bibitem[\protect\citeauthoryear{Tessendorf}{Tessendorf}{2015}]%
        {tessendorf2015advection}
\bibfield{author}{\bibinfo{person}{Jerry Tessendorf}.} \bibinfo{year}{2015}\natexlab{}.
\newblock \showarticletitle{Advection Solver Performance with Long Time Steps, and Strategies for Fast and Accurate Numerical Implementation}.
\newblock  (\bibinfo{year}{2015}).
\newblock


\bibitem[\protect\citeauthoryear{Tessendorf and Pelfrey}{Tessendorf and Pelfrey}{2011}]%
        {tessendorf2011characteristic}
\bibfield{author}{\bibinfo{person}{Jerry Tessendorf} {and} \bibinfo{person}{Brandon Pelfrey}.} \bibinfo{year}{2011}\natexlab{}.
\newblock \showarticletitle{The characteristic map for fast and efficient vfx fluid simulations}. In \bibinfo{booktitle}{\emph{Computer Graphics International Workshop on VFX, Computer Animation, and Stereo Movies. Ottawa, Canada}}.
\newblock


\bibitem[\protect\citeauthoryear{Virtanen, Gommers, Oliphant, Haberland, Reddy, Cournapeau, Burovski, Peterson, Weckesser, Bright, {van der Walt}, Brett, Wilson, Millman, Mayorov, Nelson, Jones, Kern, Larson, Carey, Polat, Feng, Moore, {VanderPlas}, Laxalde, Perktold, Cimrman, Henriksen, Quintero, Harris, Archibald, Ribeiro, Pedregosa, {van Mulbregt}, and {SciPy 1.0 Contributors}}{Virtanen et~al\mbox{.}}{2020}]%
        {scipy2020}
\bibfield{author}{\bibinfo{person}{Pauli Virtanen}, \bibinfo{person}{Ralf Gommers}, \bibinfo{person}{Travis~E. Oliphant}, \bibinfo{person}{Matt Haberland}, \bibinfo{person}{Tyler Reddy}, \bibinfo{person}{David Cournapeau}, \bibinfo{person}{Evgeni Burovski}, \bibinfo{person}{Pearu Peterson}, \bibinfo{person}{Warren Weckesser}, \bibinfo{person}{Jonathan Bright}, \bibinfo{person}{St{\'e}fan~J. {van der Walt}}, \bibinfo{person}{Matthew Brett}, \bibinfo{person}{Joshua Wilson}, \bibinfo{person}{K.~Jarrod Millman}, \bibinfo{person}{Nikolay Mayorov}, \bibinfo{person}{Andrew R.~J. Nelson}, \bibinfo{person}{Eric Jones}, \bibinfo{person}{Robert Kern}, \bibinfo{person}{Eric Larson}, \bibinfo{person}{C~J Carey}, \bibinfo{person}{{\.I}lhan Polat}, \bibinfo{person}{Yu Feng}, \bibinfo{person}{Eric~W. Moore}, \bibinfo{person}{Jake {VanderPlas}}, \bibinfo{person}{Denis Laxalde}, \bibinfo{person}{Josef Perktold}, \bibinfo{person}{Robert Cimrman}, \bibinfo{person}{Ian Henriksen}, \bibinfo{person}{E.~A. Quintero},
  \bibinfo{person}{Charles~R. Harris}, \bibinfo{person}{Anne~M. Archibald}, \bibinfo{person}{Ant{\^o}nio~H. Ribeiro}, \bibinfo{person}{Fabian Pedregosa}, \bibinfo{person}{Paul {van Mulbregt}}, {and} \bibinfo{person}{{SciPy 1.0 Contributors}}.} \bibinfo{year}{2020}\natexlab{}.
\newblock \showarticletitle{{{SciPy} 1.0: Fundamental Algorithms for Scientific Computing in Python}}.
\newblock \bibinfo{journal}{\emph{Nature Methods}}  \bibinfo{volume}{17} (\bibinfo{year}{2020}), \bibinfo{pages}{261--272}.
\newblock
\urldef\tempurl%
\url{https://doi.org/10.1038/s41592-019-0686-2}
\showDOI{\tempurl}


\bibitem[\protect\citeauthoryear{Weinan and Liu}{Weinan and Liu}{2003}]%
        {weinan2003gauge}
\bibfield{author}{\bibinfo{person}{E Weinan} {and} \bibinfo{person}{Jian-Guo Liu}.} \bibinfo{year}{2003}\natexlab{}.
\newblock \showarticletitle{Gauge method for viscous incompressible flows}.
\newblock \bibinfo{journal}{\emph{Communications in Mathematical Sciences}} \bibinfo{volume}{1}, \bibinfo{number}{2} (\bibinfo{year}{2003}), \bibinfo{pages}{317--332}.
\newblock


\bibitem[\protect\citeauthoryear{Wiggert and Wylie}{Wiggert and Wylie}{1976}]%
        {wiggert1976numerical}
\bibfield{author}{\bibinfo{person}{DC Wiggert} {and} \bibinfo{person}{EB Wylie}.} \bibinfo{year}{1976}\natexlab{}.
\newblock \showarticletitle{Numerical predictions of two-dimensional transient groundwater flow by the method of characteristics}.
\newblock \bibinfo{journal}{\emph{Water Resources Research}} \bibinfo{volume}{12}, \bibinfo{number}{5} (\bibinfo{year}{1976}), \bibinfo{pages}{971--977}.
\newblock


\bibitem[\protect\citeauthoryear{Xiong, Wang, Wang, and Zhu}{Xiong et~al\mbox{.}}{2022}]%
        {Xiong2022Clebsch}
\bibfield{author}{\bibinfo{person}{S. Xiong}, \bibinfo{person}{Z. Wang}, \bibinfo{person}{M. Wang}, {and} \bibinfo{person}{B. Zhu}.} \bibinfo{year}{2022}\natexlab{}.
\newblock \showarticletitle{{A Clebsch method for free-surface vortical flow simulation}}.
\newblock \bibinfo{journal}{\emph{ACM Trans. Graph.}} \bibinfo{volume}{41}, \bibinfo{number}{4} (\bibinfo{year}{2022}).
\newblock


\bibitem[\protect\citeauthoryear{Yang, Xiong, Zhang, Feng, Liu, and Zhu}{Yang et~al\mbox{.}}{2021}]%
        {yang2021clebsch}
\bibfield{author}{\bibinfo{person}{Shuqi Yang}, \bibinfo{person}{Shiying Xiong}, \bibinfo{person}{Yaorui Zhang}, \bibinfo{person}{Fan Feng}, \bibinfo{person}{Jinyuan Liu}, {and} \bibinfo{person}{Bo Zhu}.} \bibinfo{year}{2021}\natexlab{}.
\newblock \showarticletitle{Clebsch gauge fluid}.
\newblock \bibinfo{journal}{\emph{ACM Transactions on Graphics (TOG)}} \bibinfo{volume}{40}, \bibinfo{number}{4} (\bibinfo{year}{2021}), \bibinfo{pages}{1--11}.
\newblock


\bibitem[\protect\citeauthoryear{Yin, Mercier, Yadav, Schneider, and Nave}{Yin et~al\mbox{.}}{2021}]%
        {yin2021characteristic}
\bibfield{author}{\bibinfo{person}{Xi-Yuan Yin}, \bibinfo{person}{Olivier Mercier}, \bibinfo{person}{Badal Yadav}, \bibinfo{person}{Kai Schneider}, {and} \bibinfo{person}{Jean-Christophe Nave}.} \bibinfo{year}{2021}\natexlab{}.
\newblock \showarticletitle{A Characteristic Mapping method for the two-dimensional incompressible Euler equations}.
\newblock \bibinfo{journal}{\emph{J. Comput. Phys.}}  \bibinfo{volume}{424} (\bibinfo{year}{2021}), \bibinfo{pages}{109781}.
\newblock


\bibitem[\protect\citeauthoryear{Yin, Schneider, and Nave}{Yin et~al\mbox{.}}{2023}]%
        {yin2023characteristic}
\bibfield{author}{\bibinfo{person}{Xi-Yuan Yin}, \bibinfo{person}{Kai Schneider}, {and} \bibinfo{person}{Jean-Christophe Nave}.} \bibinfo{year}{2023}\natexlab{}.
\newblock \showarticletitle{A Characteristic Mapping Method for the three-dimensional incompressible Euler equations}.
\newblock \bibinfo{journal}{\emph{J. Comput. Phys.}} (\bibinfo{year}{2023}), \bibinfo{pages}{111876}.
\newblock


\bibitem[\protect\citeauthoryear{Zhai, Hou, Qin, and Hao}{Zhai et~al\mbox{.}}{2018}]%
        {zhai2018fluid}
\bibfield{author}{\bibinfo{person}{Xiao Zhai}, \bibinfo{person}{Fei Hou}, \bibinfo{person}{Hong Qin}, {and} \bibinfo{person}{Aimin Hao}.} \bibinfo{year}{2018}\natexlab{}.
\newblock \showarticletitle{Fluid simulation with adaptive staggered power particles on {gpus}}.
\newblock \bibinfo{journal}{\emph{IEEE Transactions on Visualization and Computer Graphics}} \bibinfo{volume}{26}, \bibinfo{number}{6} (\bibinfo{year}{2018}), \bibinfo{pages}{2234--2246}.
\newblock


\end{thebibliography}


\pagebreak

\begin{figure*}
    \includegraphics[width=0.95\linewidth]{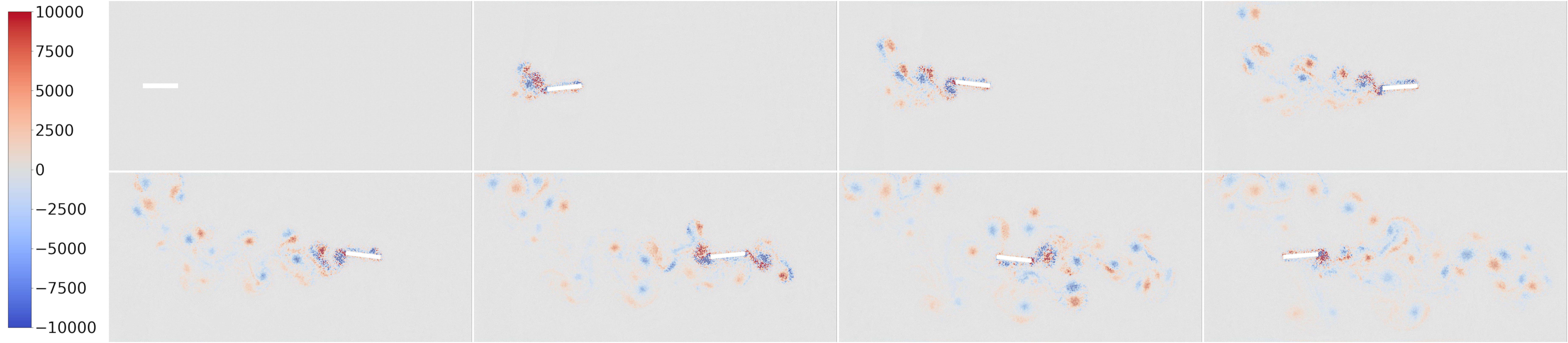}
    \vspace{-0.25cm}
    \caption{A moving and rotating board in 2D traverses the domain from left to right, and then back.}
    \label{fig:board_2d}
    \vspace{-0.0cm}
\end{figure*}

\begin{figure*}[ht]
\begin{minipage}[ht]{0.45\textwidth}    
    \centering
    \includegraphics[width=\columnwidth]{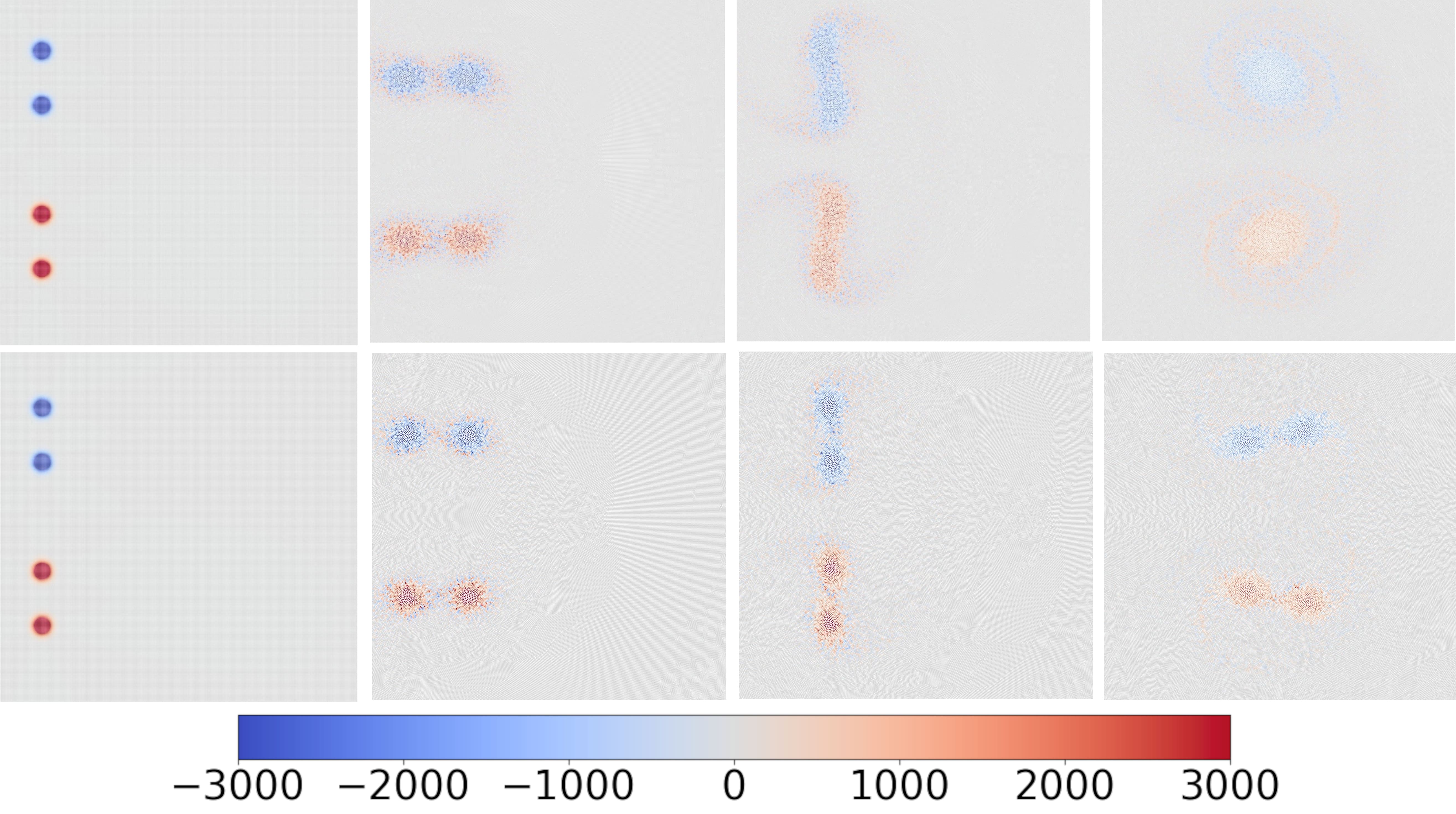}
    \caption{Leapfrog vortices in 2D. PPM (\textit{top}), our approach (\textit{bottom}).}
    \label{fig:leapfrog_2d}
\end{minipage}
\begin{minipage}[ht]{0.5\textwidth}
 \centering
  \includegraphics[width=0.45\columnwidth]{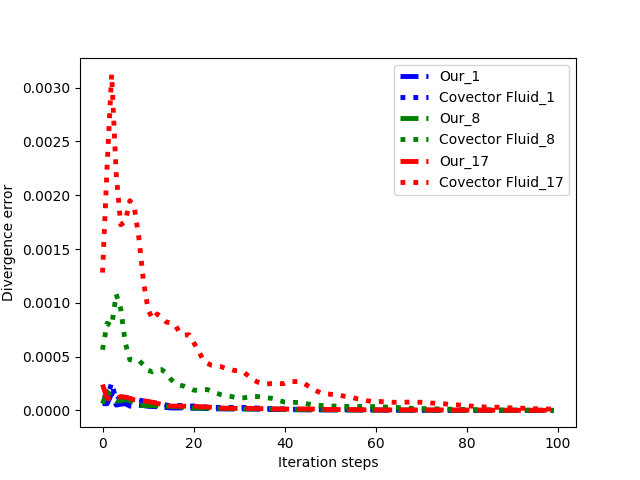}
 \vspace{-0.35cm}
 \caption{ 
 The performance when using different flow map lengths is shown in the case of both the original Covector Fluid technique, and our method, where previous pressure integration is subtracted before projection. \textit{Blue, green}, and \textit{red} lines depict divergence error-interative step curve for conjugate gradient \color{black}solver \color{black}for poisson \color{black}equation \color{black}of flow map length 1, 8, and 17, respectively. We observe an improved rate of convergence by our method.  \color{black}This plot corresponds to Leapfrog example and convergence error is calculated by averaging $|\nabla\cdot \mathbf{u}|$ on particles.\color{black}
 }
 \label{fig:ablation_convergence}
\end{minipage}
\end{figure*}

\begin{figure*}
    \vspace{-0.30cm}
    \centering
    \begin{minipage}{0.5\textwidth}
        \centering
        \includegraphics[width=0.95\textwidth]{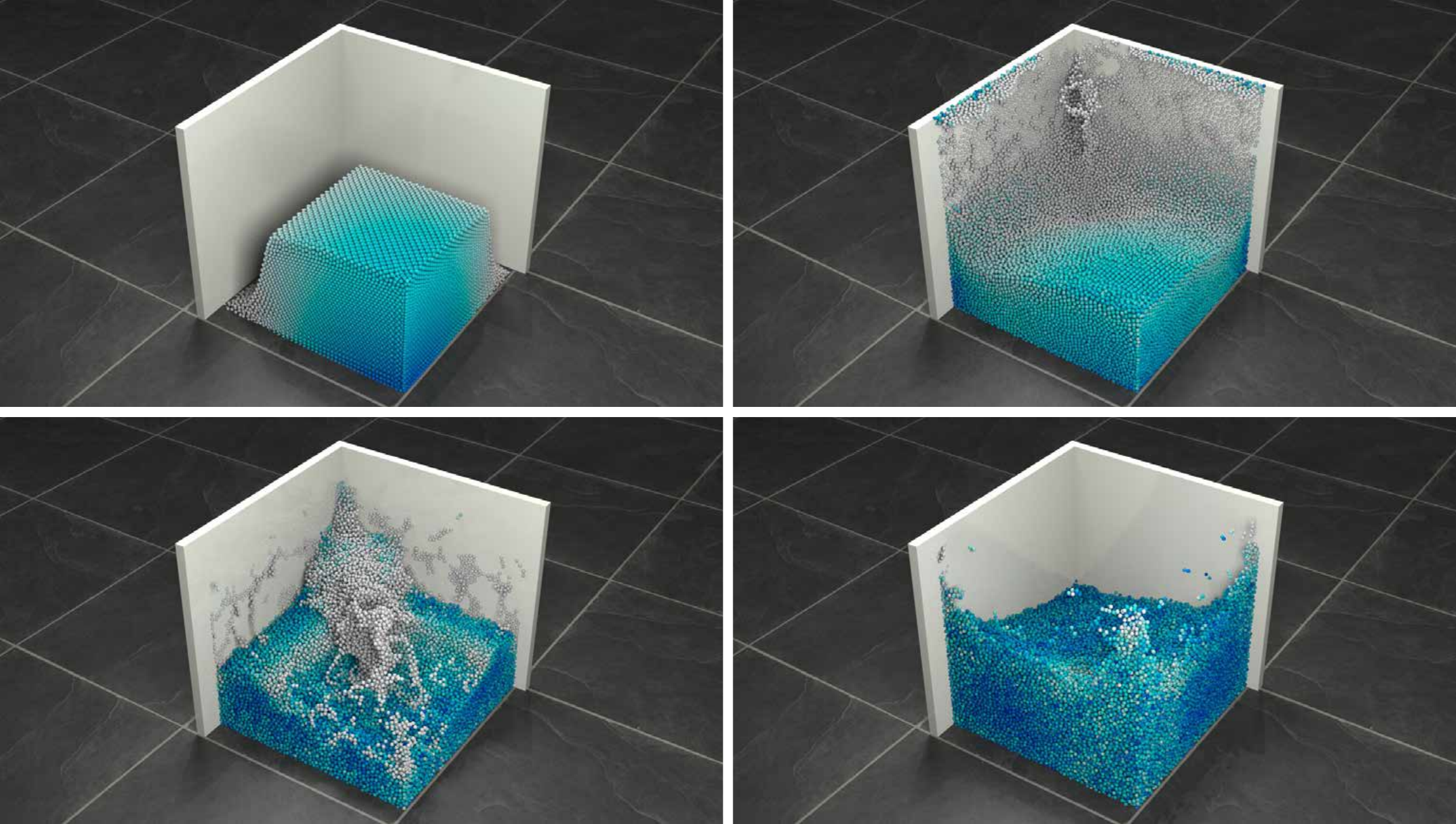}
        \vspace{-0.25cm}
        \caption{Dam break (particle visualization).}
        \label{fig:dam_break}
    \end{minipage}\hfill
    \begin{minipage}{0.5\textwidth}
        \centering
        \includegraphics[width=0.95\textwidth]{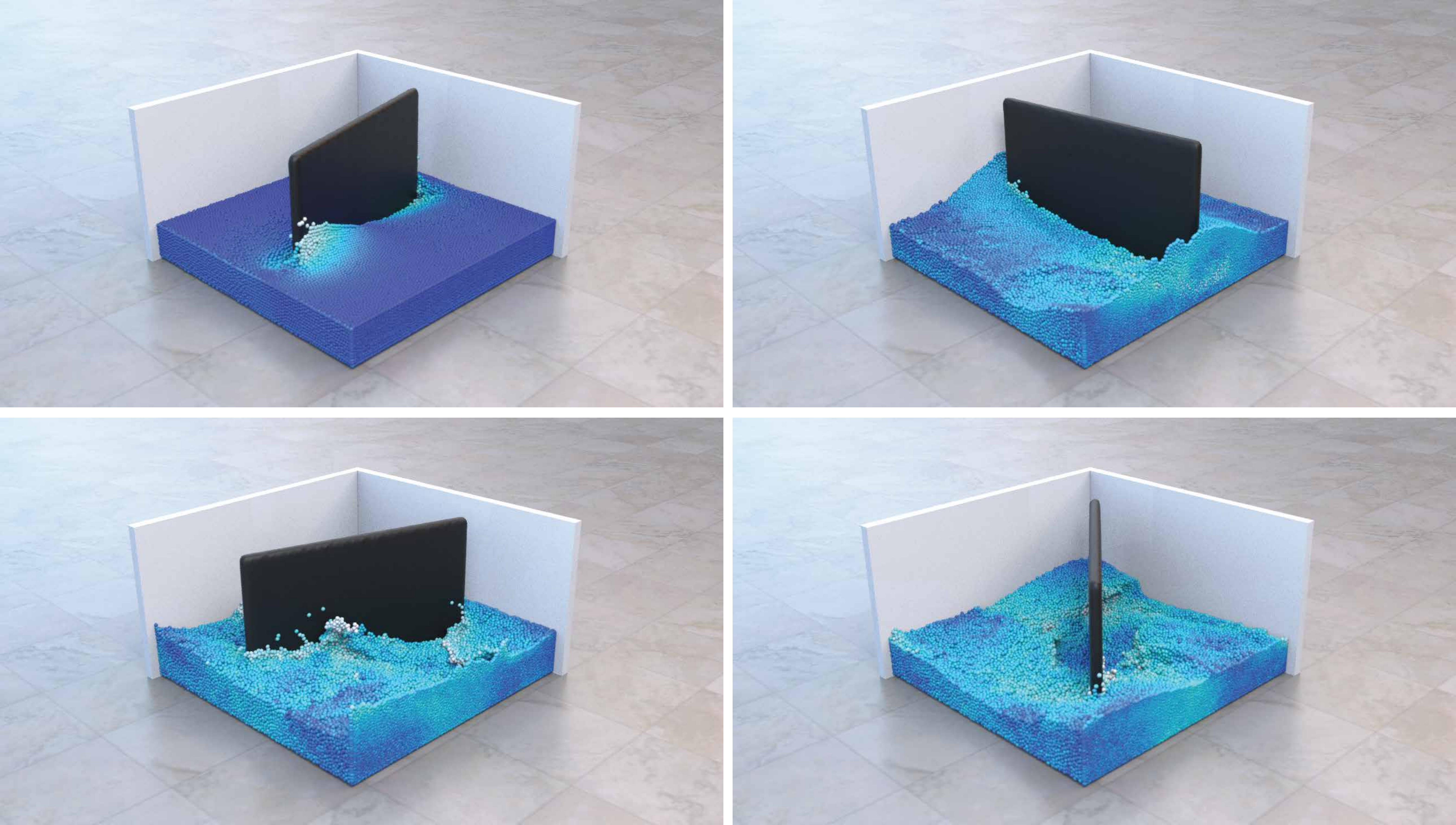}
        \vspace{-0.25cm}
        \caption{Rotating board.}
        \label{fig:board_rotate}
    \end{minipage}
\end{figure*}

\begin{figure*}
   \vspace{-0.25cm}
\begin{minipage}[t]{0.50\textwidth}
    \hspace{-0.25cm}
    \begin{subfigure}[t]{.28\columnwidth}
    \centering
    \includegraphics[width=\linewidth]{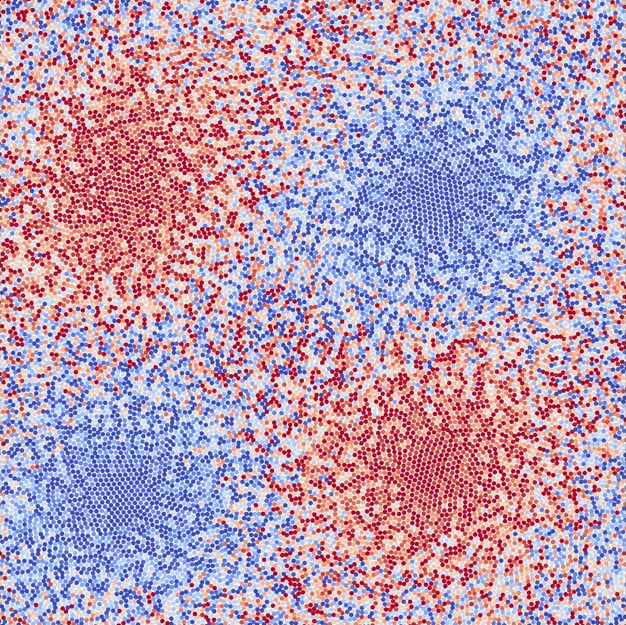}
    \caption{PPM}
    \label{fig:sub1}
  \end{subfigure}%
  \hspace{0.05cm}
  \begin{subfigure}[t]{.358\columnwidth}
    \centering
    \includegraphics[width=\linewidth]{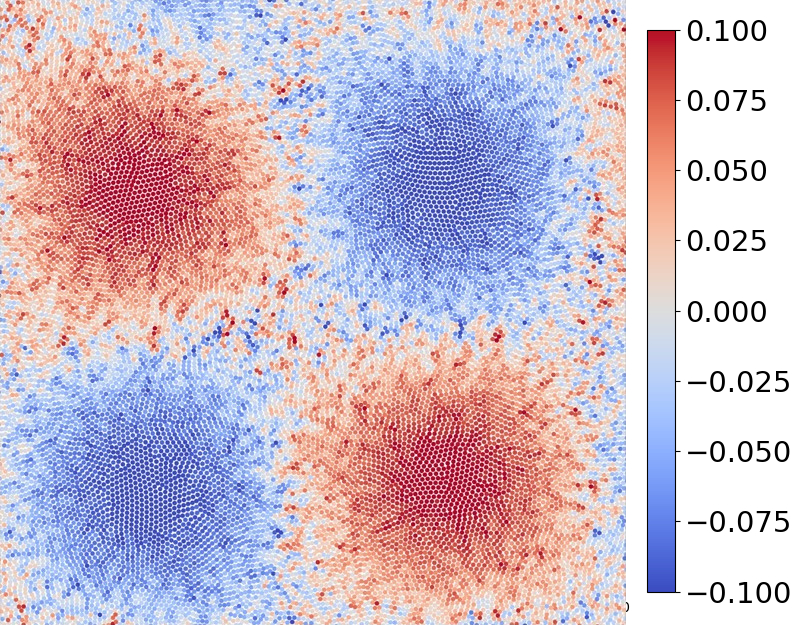}
    \caption{Our method}
    \label{fig:sub2}
  \end{subfigure}
  \begin{subfigure}[t]{.33\columnwidth}
    \centering
    \includegraphics[width=0.9\linewidth]{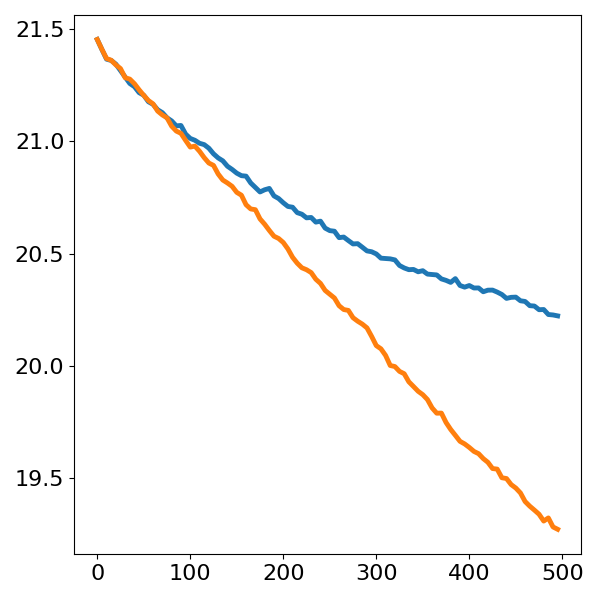}
  \end{subfigure}
  \vspace{-0.35cm}
  \caption{Qualitative and quantitative evaluation of simulating Taylor-Green vortices. (a) and (b) show the state for each method after $700$ time steps. (c) shows the volume-averaged kinematic energy (y-axis) over time steps (x-axis): the energy dissipation for PPM (orange) and our method (blue) over $500$ time steps.  \color{black}Here, coloring represents the magnitude and sign of vorticity.\color{black}
  \label{fig:taylor_green_2d}}
  \hfill
\end{minipage}
  \hspace{0.5cm}
\begin{minipage}[t]{0.4\textwidth}
    \vspace{-2.5cm}
  \begin{subfigure}[ht]{.4\columnwidth}
    \centering
    \includegraphics[width=\linewidth]{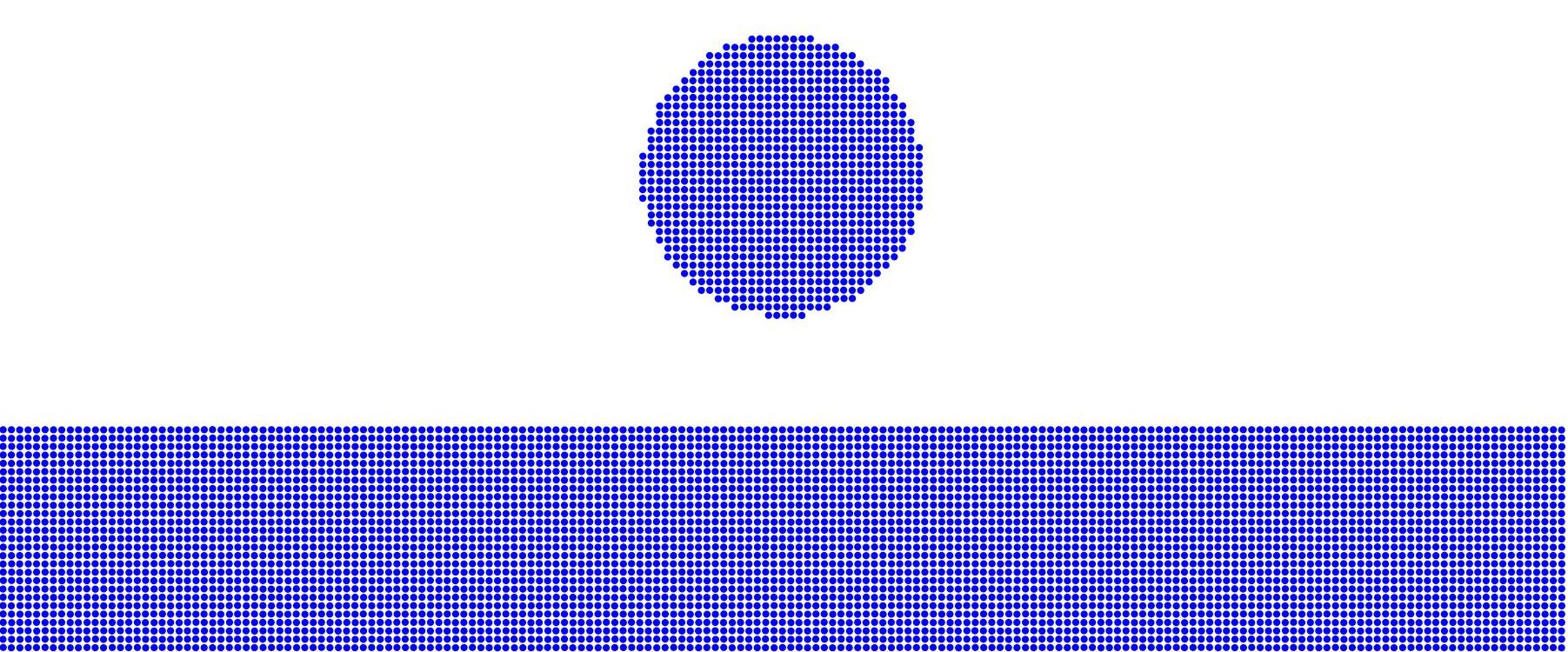}
    \label{fig:ablation_without_1}
  \end{subfigure}
  \begin{subfigure}[ht]{.4\columnwidth}
    \centering
    \includegraphics[width=\linewidth]{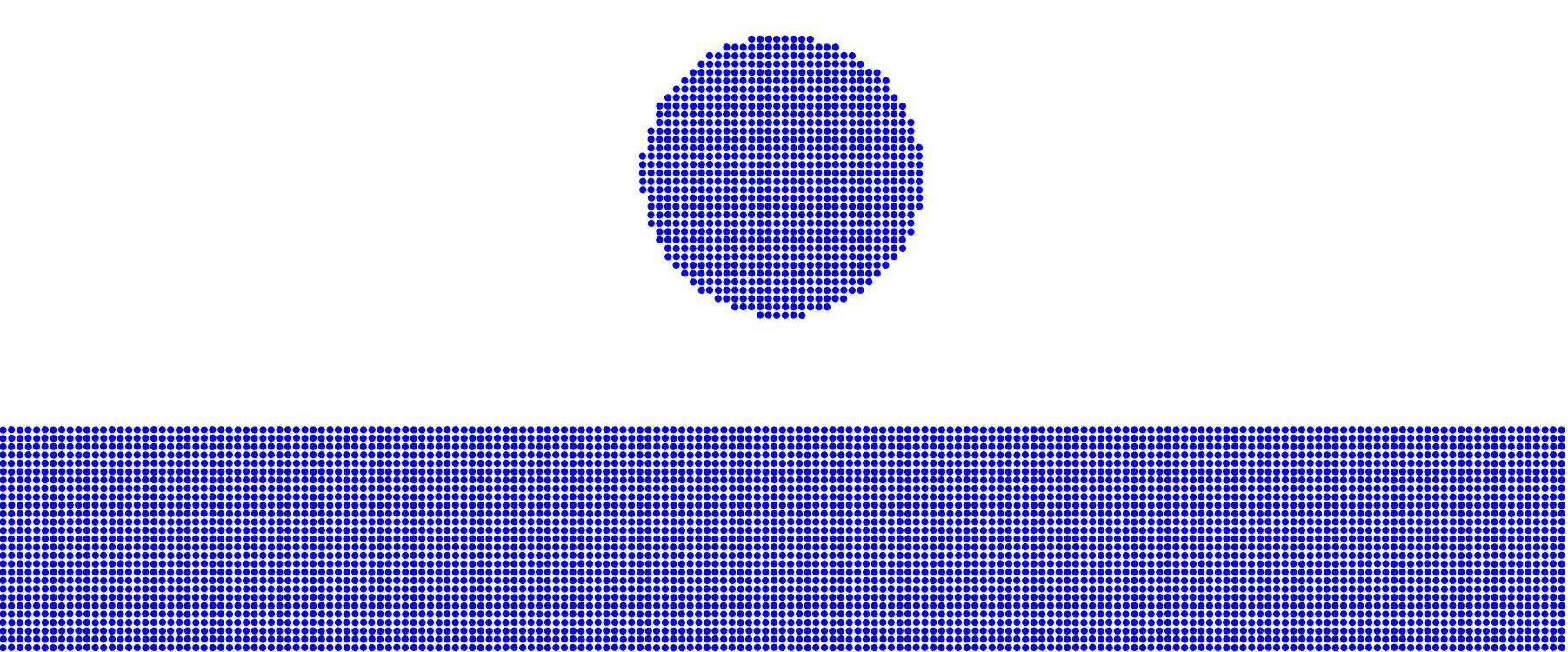}
    \label{fig:ablation_our_1}
  \end{subfigure}
  \begin{subfigure}[ht]{.4\columnwidth}
    \centering
    \includegraphics[width=\linewidth]{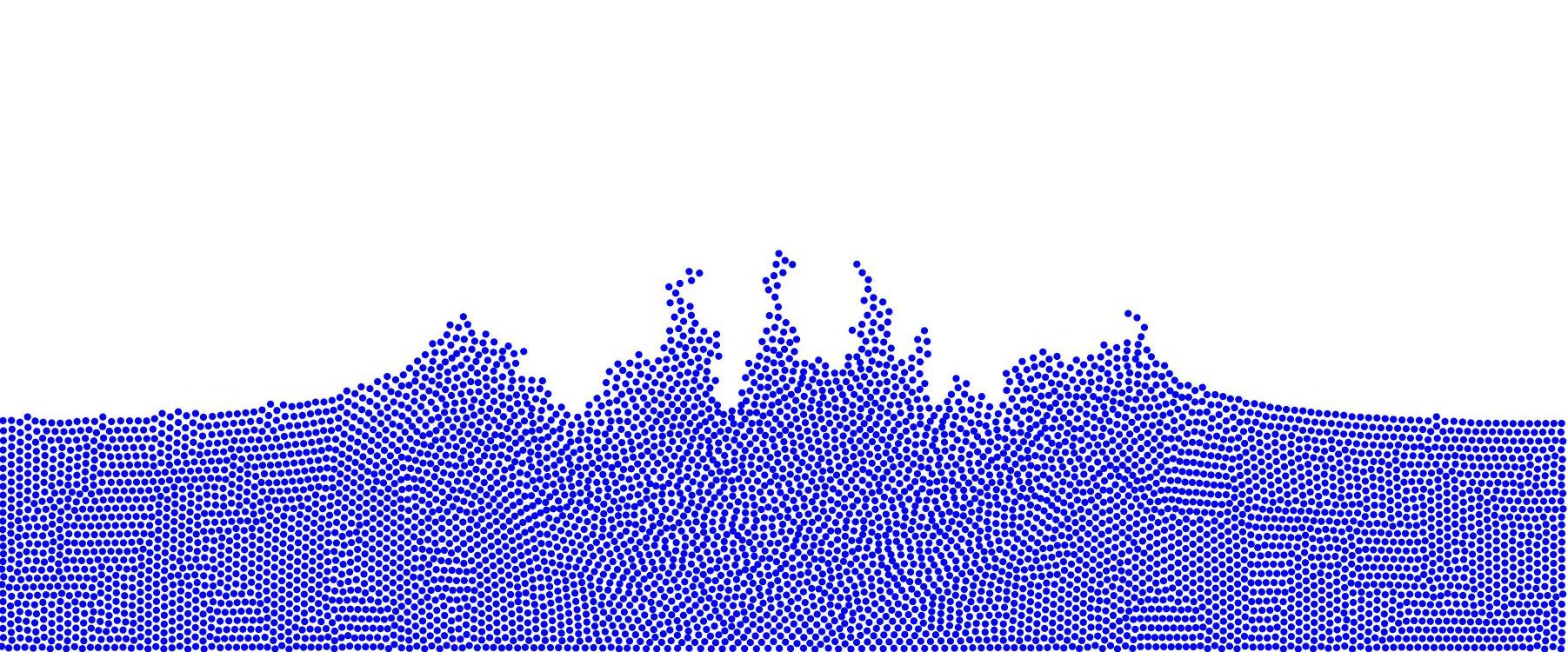}
    \caption{Without}
    \label{fig:ablation_without_7}
  \end{subfigure}
\hspace{0.5cm}
  \begin{subfigure}[ht]{.4\columnwidth}
    \centering
    \includegraphics[width=\linewidth]{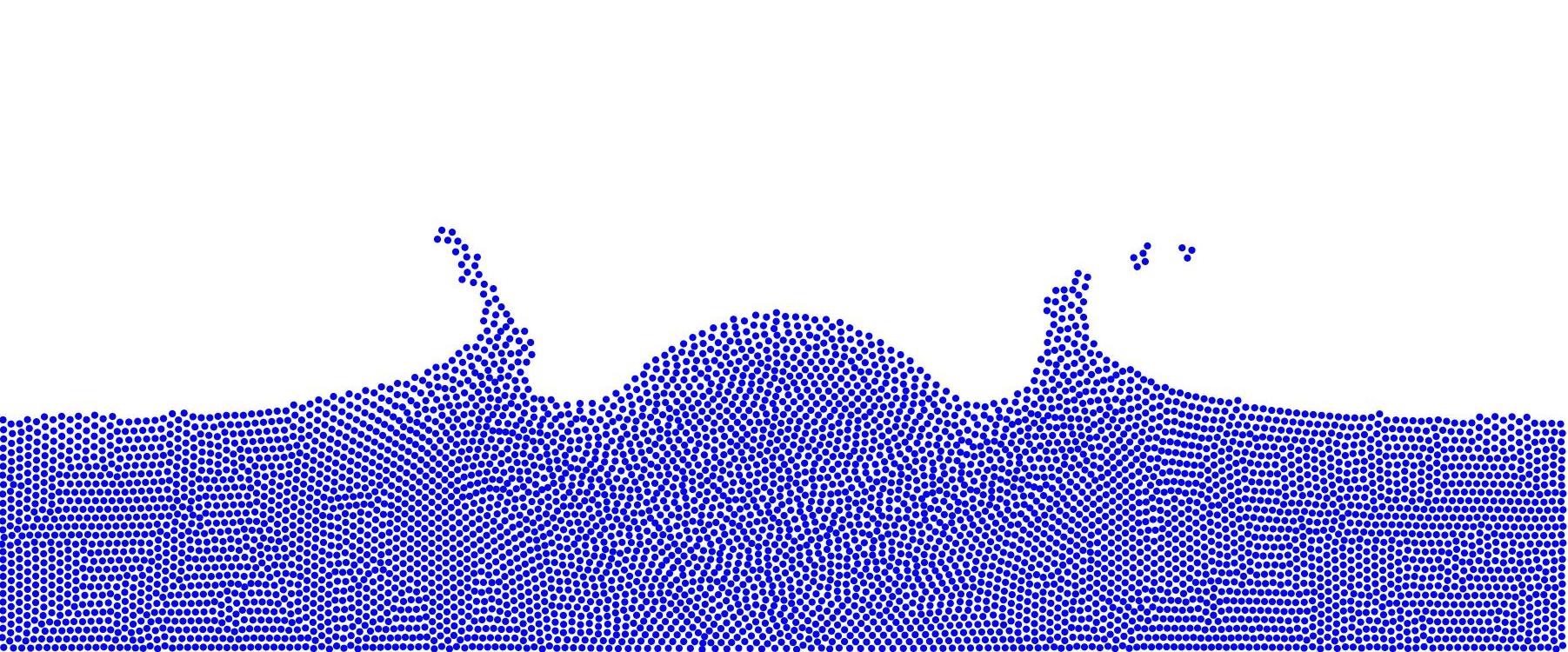}
    \caption{Our method}
    \label{fig:ablation_our_7}
  \end{subfigure}
\vspace{-0.315cm}
  \caption{Ablation study on the effect of our novel free surface treatment. We observe the same ball of fluid particles being dropped into a still body of water.When not using our method (on the left), due to inaccuracies in the $\mathcal{T}_s^r$ approximation, strange shapes appear at the free surface. When using our method (on the right), the shape of the free surface is correct.}
  \label{fig:ablation_study_free_surface}
\end{minipage}
    \vspace{-0.35cm}
\end{figure*}

\begin{figure*}
\centering
\includegraphics[width=0.9\textwidth]{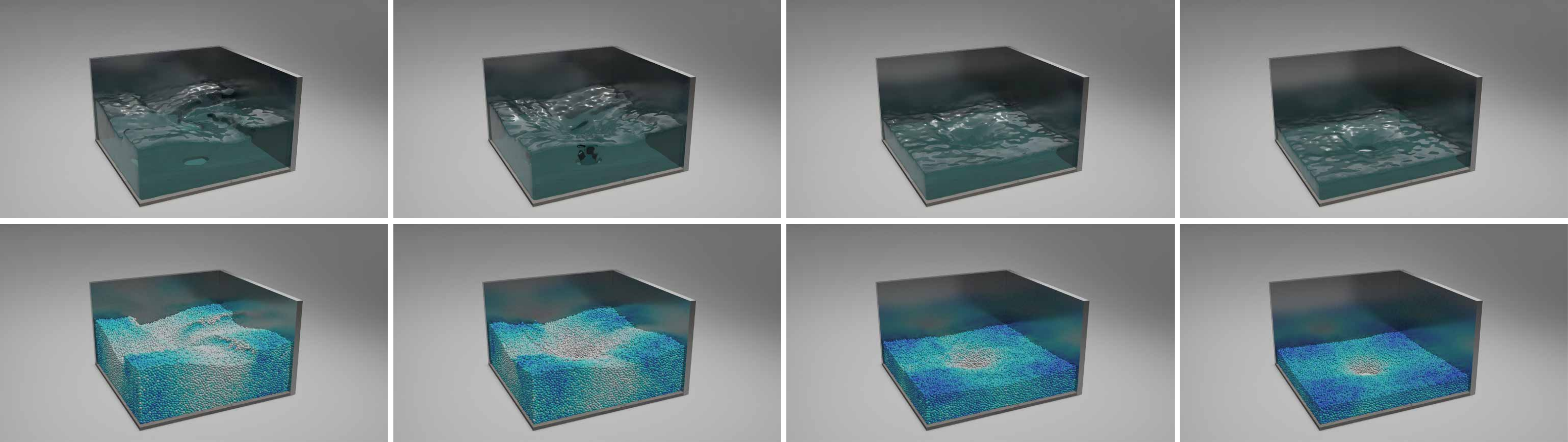}
\vspace{-0.25cm}
\caption{Single sink. Surface rendering \textit{(top)} and particle visualization \textit{(bottom)}.}
\label{fig:one_sink_both}
\end{figure*}

\begin{figure*}
\centering
\includegraphics[width=0.9\textwidth]{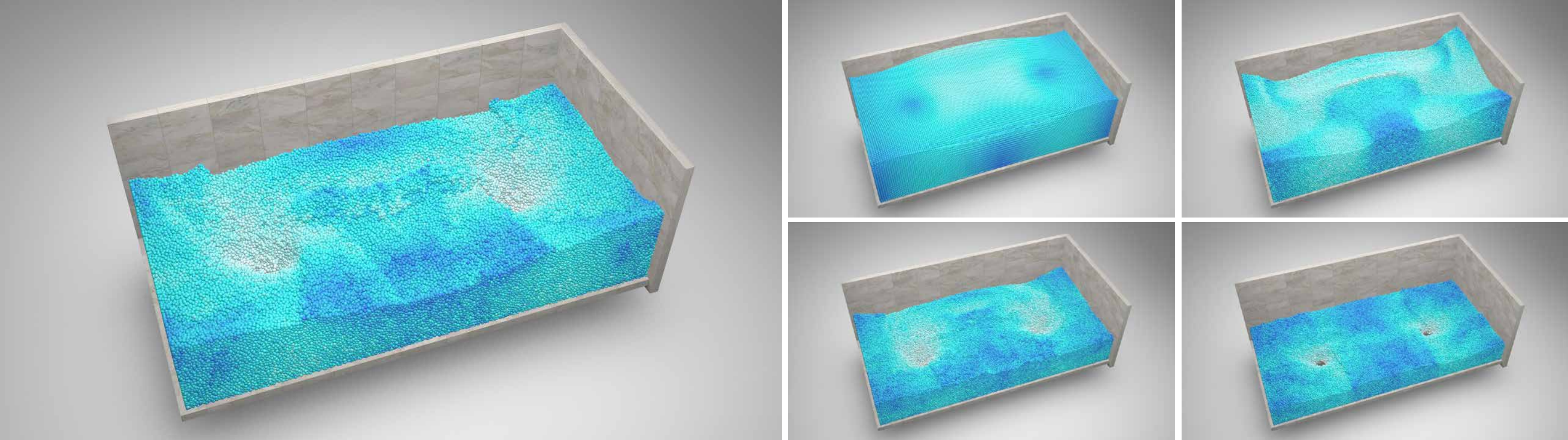}
\vspace{-0.25cm}
\caption{Double sink (particle visualization).}
\label{fig:two_sink_particles}
\end{figure*}

\begin{figure*}
\centering
\includegraphics[width=0.9\textwidth]{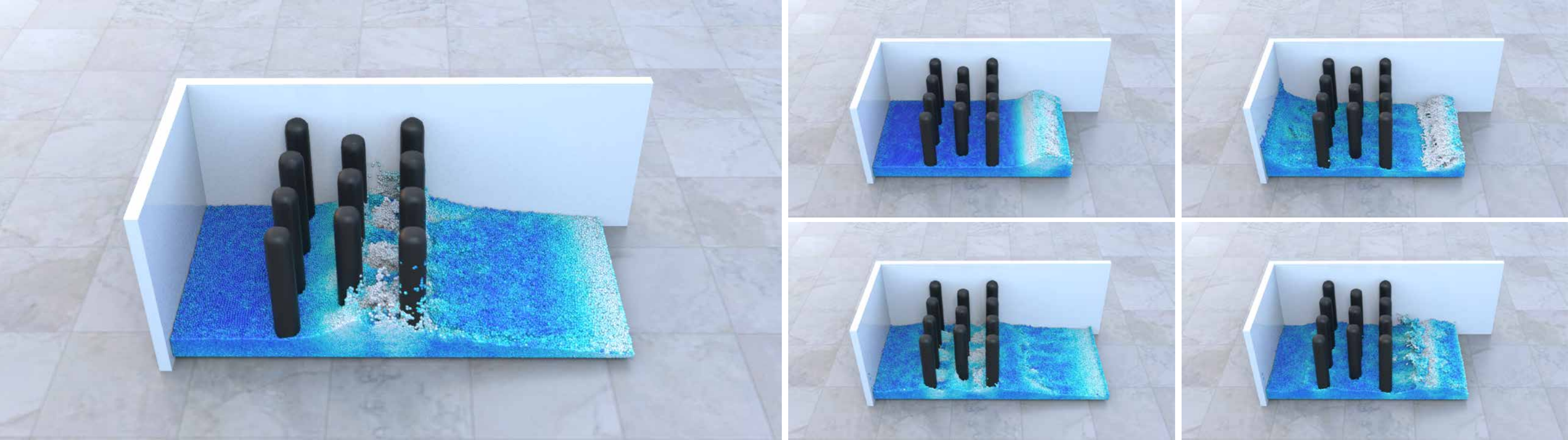}
\vspace{-0.25cm}
\caption{Waves generated in a water tank with cylindrical obstacles.}
\label{fig:wave-columns}
\end{figure*}

\begin{figure*}
\hspace{-0.5cm}
    \centering
    \begin{minipage}[t]{0.3\textwidth}
        \centering
        \includegraphics[width=5cm]{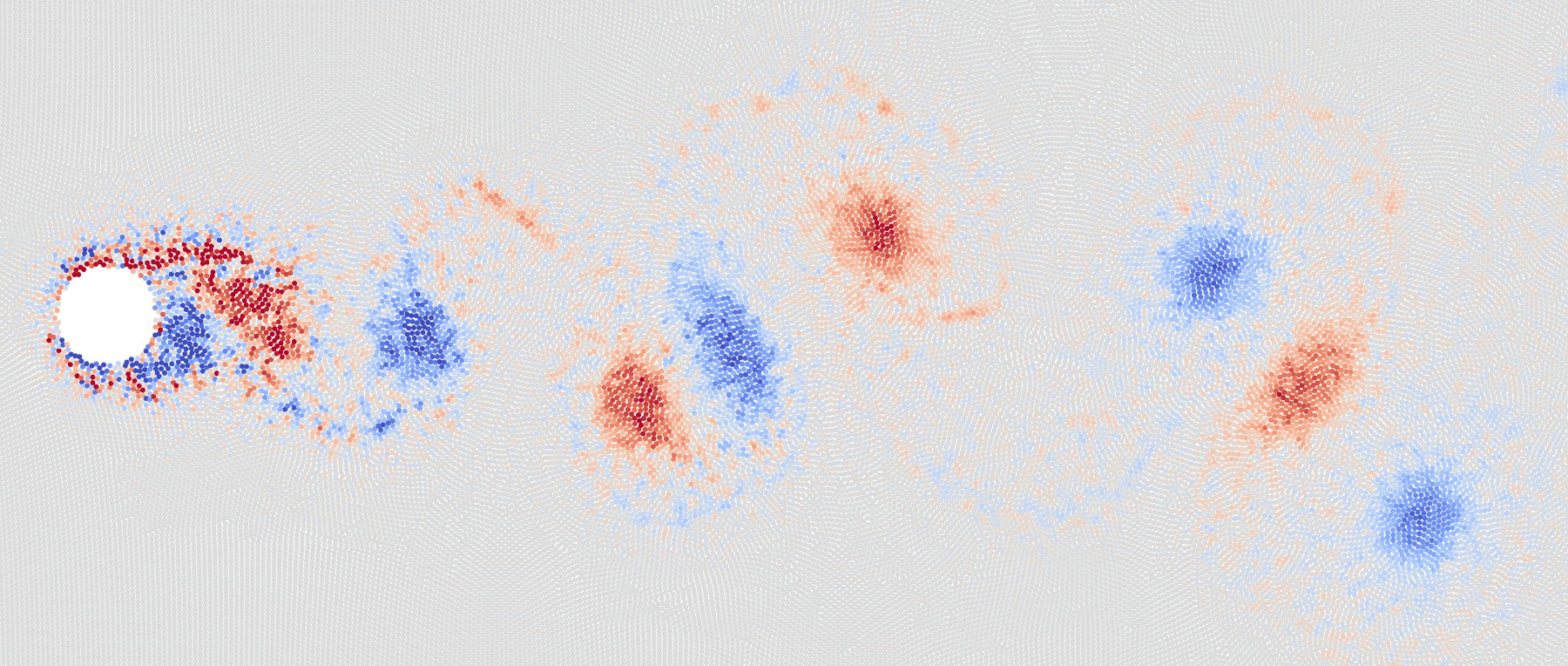}
        \vspace{-0.25cm}
        \caption{K\'arm\'an vortex street.}  \label{fig:karman}
    \hspace{0.2cm}
    \end{minipage}
    \begin{minipage}[t]{0.2\textwidth}
        \centering
        \includegraphics[width=4cm]{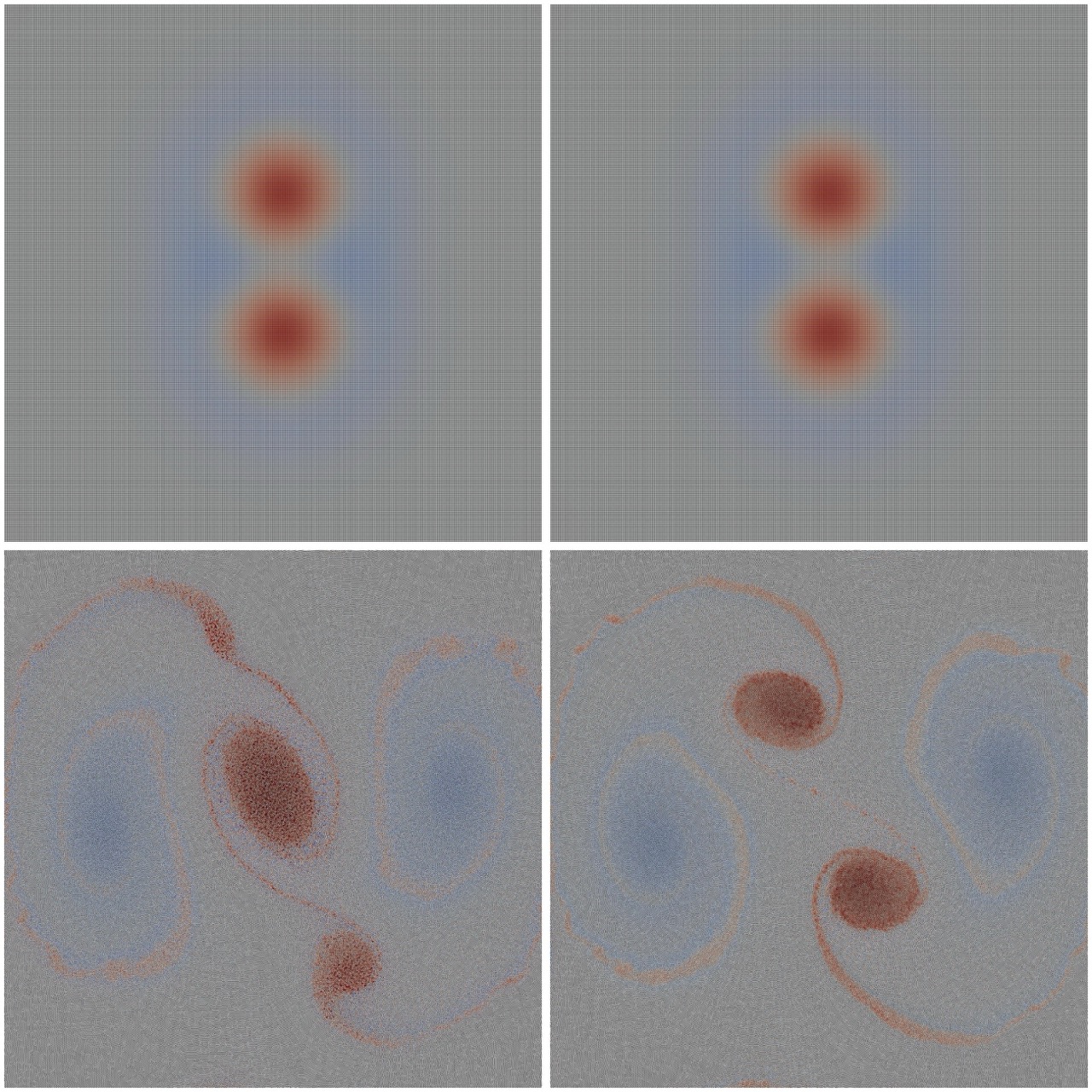}
        \vspace{-0.55cm}
        \caption{Taylor vortices. Initial state \textit{(top)} and at $300$ steps \textit{(bottom)}. PPM \textit{(left)}, Ours \textit{(right)}.}
        \label{fig:2d_taylor}
    \end{minipage}
    \begin{minipage}[t]{0.45\textwidth}
    \centering
\includegraphics[width=7cm]{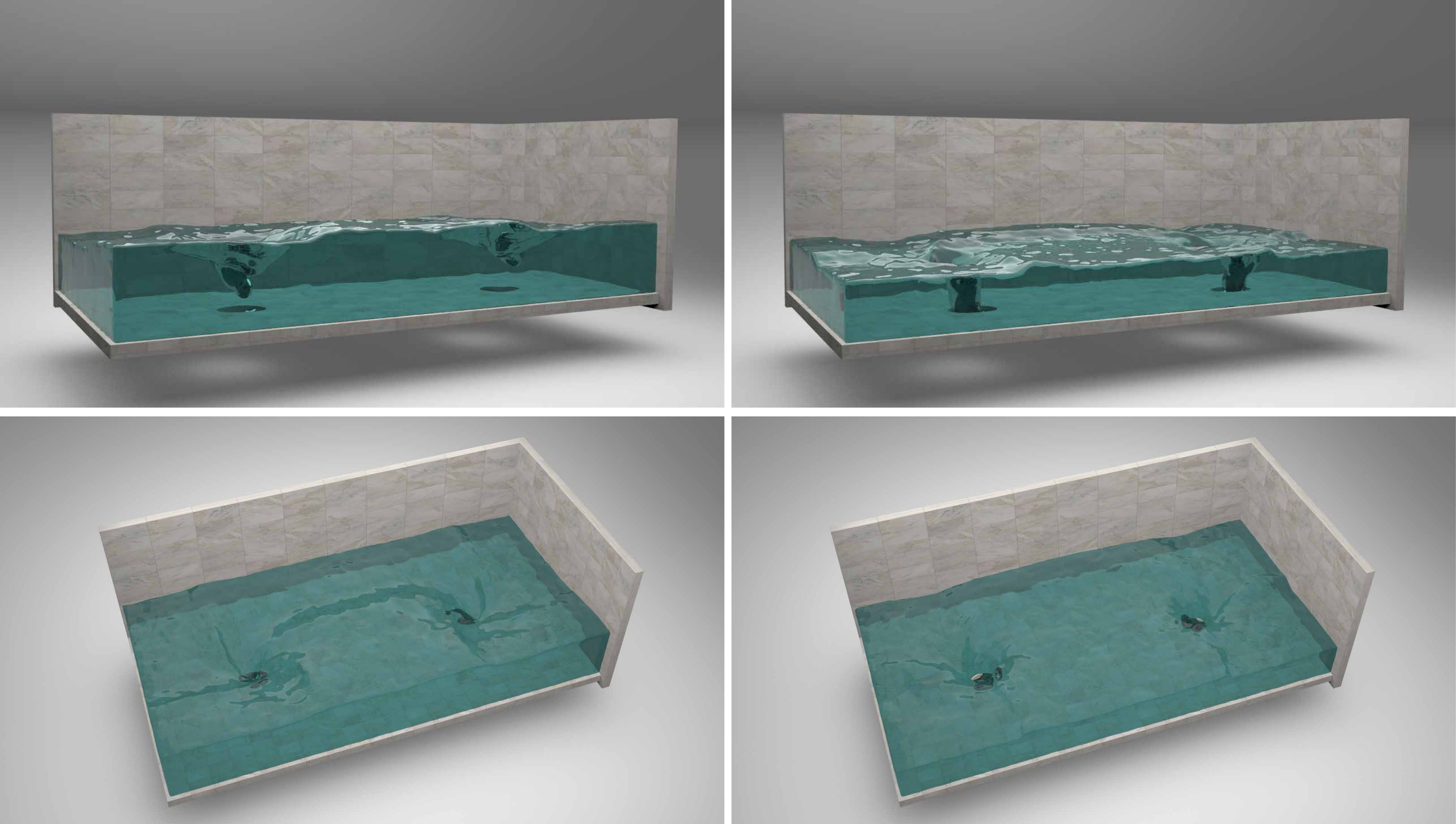}
\vspace{-0.25cm}
\caption{Double sink (surface rendering).}
    \label{fig:two_sink_surface}
\end{minipage}
\end{figure*}


\end{document}